\documentclass[aps,prd,twocolumn]{revtex4}
\usepackage{graphicx}
\usepackage{amssymb}
\usepackage{amsmath}
\usepackage{epstopdf}
\usepackage{color}
\begin{document}

\def\SU2{\ensuremath{{SU(2)}}}
\def\sdmass{\ensuremath{\sqrt{a_o}/2}}
\newcommand{\bea}{\begin{eqnarray}}
\newcommand{\eea}{\end{eqnarray}}
\def\half{\frac{1}{2}}
\def\beq{\begin{equation}}
\def\eeq{\end{equation}}
\def\beqn{\begin{eqnarray}}
\def\eeqn{\end{eqnarray}}

 \title{Generalized Uncertainty Principle
 and Self-dual Black Holes} 
\author{$^1$Bernard Carr, $^2$Leonardo Modesto, $^3$Isabeau Pr\'emont-Schwarz}
\affiliation{$^{1}$Astronomy Unit, Queen Mary University of London, Mile End Road, London E1 4NS, UK\\
$^{2}$Perimeter Institute for Theoretical Physics, 31 Caroline St.N., Waterloo, ON N2L 2Y5, Canada \\ $^{3}$Max-Planck-Institut f\"ur Gravitationsphysik,
Am M\"uhlenberg 1, Golm, D-14476 Golm, Germany}


\begin{abstract}

The Generalized Uncertainty Principle suggests corrections to the Uncertainty Principle as the energy increases towards the Planck value. It provides a natural transition between the expressions  for the Compton wavelength below the Planck mass and the black hole event horizon size above this mass. It also suggests corrections to the the event horizon size as the black hole mass falls towards the Planck value, leading to the concept of a Generalized Event Horizon. Extrapolating below the Planck mass suggests the existence of a new class of black holes, whose size is of order the Compton wavelength for their mass. Such sub-Planckian black holes have recently been discovered in the context of loop quantum gravity and it is possible that this applies more generally. 
This suggests an intriguing connection between black holes, the Uncertainty Principle and quantum gravity. 


\end{abstract}


\maketitle
\section{Introduction}

Many interesting insights can be obtained by classifying physical systems in terms of their mass-scale $M$ and length-scale $R$. For example, the location in an  $(M,R)$ diagram of a vast range of physical structures (atoms, exploding black holes, humans, planets, stars, galaxies etc.) can be understood using simple physical arguments \cite{carr-rees}.  Of particular interest in this diagram is the Compton line $R = \hbar/(Mc)$, corresponding to the Uncertainty Principle,  and the Schwarzschild line $R =2GM/c^2$, corresponding to the existence of black holes. These two lines differentiate between classical (i.e. Newtonian), quantum and relativistic  systems; this division will play an important role in the considerations of this paper.

The Compton and Schwarzschild lines intersect at the Planck scales, $M_P \sim 10^{-5}$g and $R_P \sim 10^{-33}$cm,  and presumably change their form as one approaches the intersection due to quantum gravity effects. The modification to the Compton line as $M$ increases towards $M_P$ corresponds to what is termed the Generalized Uncertainty Principle (GUP) and several calculations (involving string theory, loop quantum gravity and purely heuristic arguments) suggest that the Compton  scale has the generalized form $ R = (\hbar/Mc) [1  + \alpha (M/M_P)^2]$, where the constant $\alpha$ 
depends on the particular model \cite{adler1}. 

Although this formula is usually only applied in the limit $M \ll M_P$, it is striking that it goes over to the Schwarzschild line for $M \gg M_P$ (up to a numerical factor). The Compton wavelength is less than the Planck scale in this mass regime and, as discussed later, there are several heuristic reasons for suspecting that the Uncertainty Principle should be replaced by the black hole condition there. In particular, an external observer cannot localize an object on a scale less than its Schwarzschild radius because it cannot be seen once it is smaller than that.
Also, it is well known that quantum radiation is associated with a black hole event horizon \cite{hawking}, even though this is a long way from the Compton domain in the $(M,R)$ diagram. 
This suggests that there may be some profound connection between the Uncertainty Principle  on microscopic scales and black holes on macroscopic scales. In some sense, the quantum regime seems to {\it become} the black hole regime.  

The GUP also suggests a
modification to the Schwarzschild line as $M$ decreases towards $M_P$ of the form $R = (2GM/c^2) [ 1  + \beta (M_P/ M)^2]$ for some constant $\beta$, with the second term representing a small perturbation to the usual result for $M \gg M_P$. Indeed, one might regard this as defining a Generalized Event Horizon (GEH), which
asymptotes to the Compton form for $M \ll M_P$ just as the modified Compton scale asymptotes to the Schwarzschild form for $M \gg M_P$. 
The simplest expression which asymptotes to the Compton and Schwarzschild forms at small and large $M$ is $R = (2GM/c^2) +  (\hbar/Mc)$, so it is interesting that this naturally yields the above GUP and GEH forms to within a numerical constant. More generally, one might consider an expression of the form $R = [(2GM/c^2)^n +  (\hbar/Mc)^n]^{1/n}$  for some integer $n$.

At first sight it makes no sense to consider black holes with $M \ll M_P$
because they would be smaller than their own Compton wavelength. However, it has recently been discovered that loop quantum gravity (LQG) permits the existence of a new type of black hole which is contained within a wormhole linking two asymptotic regions \cite{Modesto:2008im}. In one region it has the usual mass $M \gg M_P$ but in the other region it has the ``dual'' mass $M_P^2/M \ll M_P$. The sub-Planckian black hole has the Compton scale for mass $M$ rather than the (much smaller) Schwarzschild scale, just as predicted by the GEH expresssion. Although this is not the usual type of black hole, since it is hidden behind the throat of a wormhole, this supports the proposed connection between black holes and the Uncertainty Principle. However, it corresponds to $n=2$ rather than $n=1$.

The notion that there is some generalized expression for the Compton and Schwarzschild scales which unifies them 
is here described as the Black Hole Uncertainty Principle (BHUP) correspondence. The crucial implication of this correspondence is that the functional dependence of the black hole radius on $M$ directly relates to the functional dependence of the uncertainty in position $\Delta x$ on the uncertainty in momentum $\Delta p$. Although this dependence has a very specific form in LQG, we speculate that this idea could also apply in other approaches to quantum gravity, including higher dimensional ones. However, for reasons discussed later, the BHUP correspondence does not seem to apply in string theory.

A particularly interesting application of the BHUP correspondence concerns the link with black hole thermodynamics. 
A perspective in which the black hole boundary {\it becomes} the quantum boundary in the macroscopic domain naturally accommodates Hawking's prediction of 
black hole radiation.
But it also goes beyond the Hawking result in predicting how the temperature should be modified below the Planck mass. Specifically, it suggests that the temperature should {\it decrease} as $M$ falls below $M_P$, giving rise to 
sub-Planckian relics which are effectively stable (at least on a cosmological timecale). While the possibility of black hole relics did not apply in the original Hawking calculation, it does arise in  various other scenarios \cite{bowick}.

\if
One approach to quantum gravity, Loop Quantum Gravity
(LQG) \cite{LQGgeneral,LQGgeneral2,LQGgeneral3}, has given rise to
models that describe the very early universe. This simplified
framework, which uses a minisuperspace approximation, has been shown
to resolve the initial singularity problem \cite{Bojowald}.
 A black hole metric in this model, known as the loop black hole (LBH)
 \cite{Modesto:2008im}, has  a property of self-duality
that removes the singularity and replaces it with
another asymptotically flat region. The
thermodynamic properties of these self-dual black holes have been
examined in \cite{Modesto:2008im,poly}, and
in \cite{Hossenfelder:2009fc} the dynamical aspects of the collapse and
evaporation were studied. The black hole space-time has also been studied in a midi-superspace reduction of LQG \cite{GP}.
\fi

This paper is organized as follows. Sec.~II considers the standard expressions for the Compton and Schwarzschild lines and explains how these divide the ($M,R$) diagram into the quantum, classical, relativistic and quantum gravity regimes. Sec.~III discusses the modifications required as $M \rightarrow M_P$ from {\it below}, corresponding to the Generalized Uncertainty Principle, while Sec.~IV discusses the modifications required as $M \rightarrow M_P$ from {\it above}, corresponding to the Generalized Event Horizon. Sec.~V shows how the BHUP correspondence
elucidate the  nature of black hole quantum emission and suggests how the black hole temperature is modified for $M \ll M_P$. 
Sec.~VI reviews some properties of the sub-Planckian loop black hole solutions and
links these solutions to the BHUP correspondence. Some final conclusions are drawn in Sec.~VII.

 \section{Compton versus Schwarzschild}
 
The Heisenberg Uncertainty Principle (HUP) implies that the uncertainty in the position and momentum of a particle must satisfy \cite{heisenberg}
 \begin{equation}
 \Delta x > \hbar/ (2 \Delta p) \,  ,
\label{UP}
 \end{equation}
where the factor of $2$ is sometimes omitted but must be included if one interprets  the uncertainties as root-mean-squares \cite{kennard}. This result can be obtained, for example, by considering the momentum imparted to a particle by the photon used to determine its position.  
It can also be derived from the commutation relation between the position and momentum operators.
For any two operators
$A $ and $B$, with commutator $[A,B] = AB-BA$, one has
\begin{equation}
\sigma_A \sigma_B > \frac{1}{2} | \left< [A,B] \right>|  
\end {equation}
where  $\sigma_A= \sqrt{\left<A^2 \right> - \left<A \right>^2}$ and $\left< A \right > = \left <\psi |A|\psi \right>$ etc.
The HUP derives from the relation $[x,p] = i \hbar$.

Since the momentum of a particle of mass $M$ is bounded by 
$Mc$, an immediate implication of the uncertainty limit is that  one cannot localize a particle on a scale less than its Compton wavelength,
 \begin{equation}
R_C = \hbar/(Mc) \, . 
\label{compton}
 \end{equation}
The factor of $2$ has now been dropped but $R_C$ is sometimes taken to be $2\pi$ times larger than this, corresponding to using $h$ rather than $\hbar$ in Eq.~\eqref{compton}. Although $M$ is usually interpreted as the rest mass, this formula applies in any relativistic frame, since $M$ is increased and $R_C$ is decreased by a Lorentz factor. For a zero-rest-mass particle like a photon, $M$ must interpreted in terms of the total energy
and $R_C$ is just the photon wavelength.

\begin{figure}
 \begin{center}
 \includegraphics[height=5.5cm]{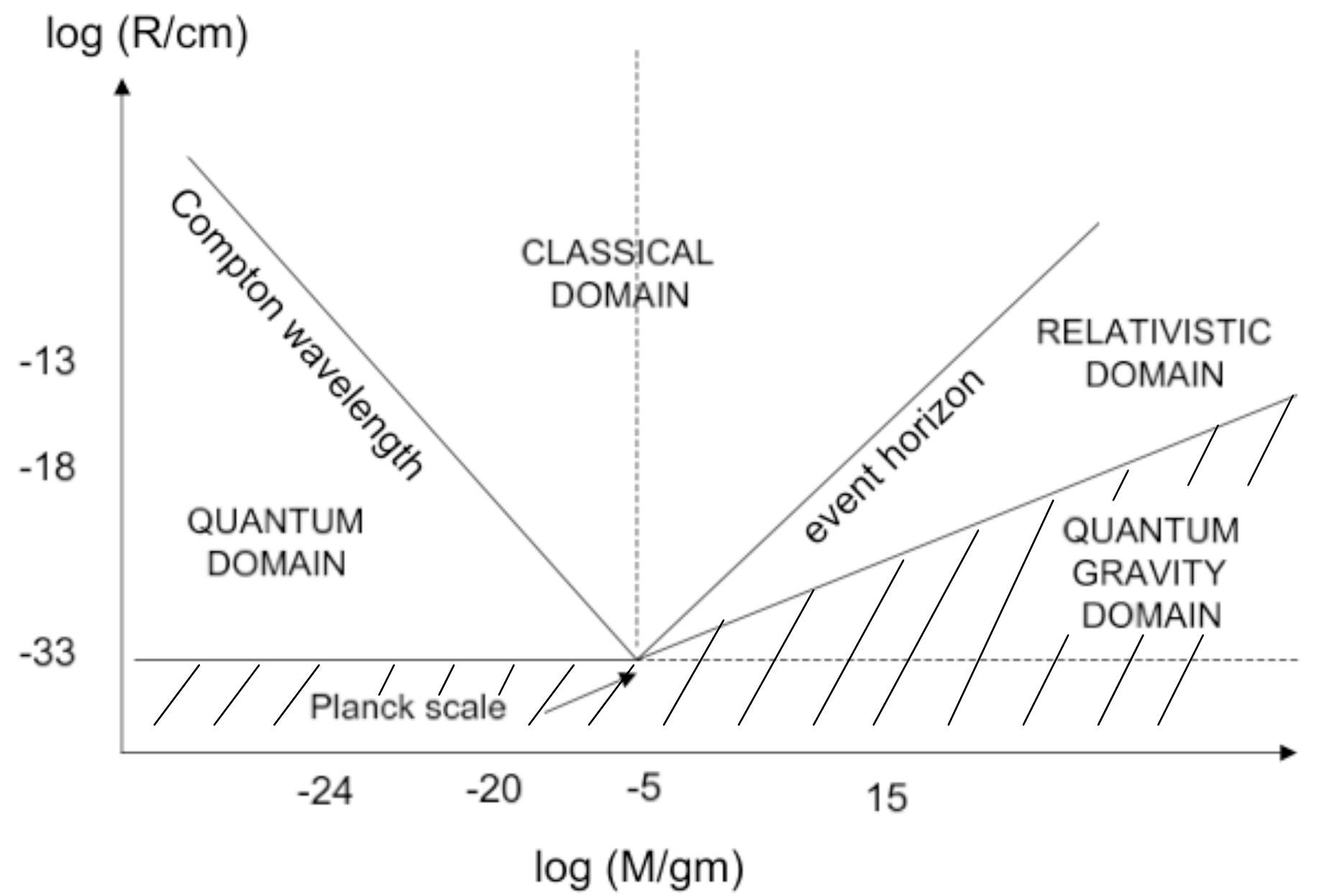}
   \end{center}
  \caption{ \label{MR} Showing the division of the ($M,R$) diagram into the classical, quantum, relativistic and quantum gravity domains. The boundaries are specified by the Compton wavelength, the Schwarzschild radius and the Planck density. The GUP smoothes out the transition where the classical and quantum boundaries intersect. }
  \end{figure}

In the $(M,R)$ diagram of Fig.~\ref{MR}, the region corresponding to $R<R_C$ 
might  be regarded as the ``quantum domain'', in the sense that the classical description breaks down there. 
This region is not unphysical but it cannot be occupied by single-particle states because
pairs of particle of mass $M$ are produced whenever the mass is compressed within the scale $\hbar /(Mc)$, the uncertainty in the energy $\Delta E \approx c \Delta p$ then exceeding $Mc^2$.
So quantum field theory, which describes the creation and annihilation of particles, must pertain in this region. Also space behaves strangely below the Compton boundary because of  quantum entanglement and non-local effects. Recent developments suggest that these effects are intimately connected with the Uncertainty Principle \cite{oppenheim}. Although quantum theory is usually associated with microphysics, the boundary extends to arbitrarily large values of $R$.

An object of mass $M$ forms a black hole if it is compressed enough to form an event horizon.
For  a spherically symmetric object, general relativity implies that this corresponds to 
the Schwarzschild radius,
 \begin{equation}
R_S = 2GM/c^2 \, .
\label{schwarzschild}
 \end{equation}
The region $R<R_S$ might be regarded as the ``relativistic domain" and there is no stable classical configuration in this part of Fig.~\ref{MR}.  It is well known that time behaves strangely at an event horizon, in the sense that it freezes there from the perspective of an external observer, 
although it still flows for an observer who falls towards the central singularity. General relativity is often associated wth the macroscopic domain (eg. it is essential for cosmology) but we note that the black hole boundary extends down to very small values of $R$.

The boundaries given by Eqs.~\eqref{compton} and \eqref{schwarzschild} intersect at around the Planck scales,
 \begin{equation}
 R_P = \sqrt{ \hbar G/c^3} \sim 10^{-33} \mathrm {cm}, \quad 
 M_P = \sqrt{ \hbar c/G} \sim 10^{-5} \mathrm g \, ,
 \label{planck} \nonumber 
\end{equation}
and they divide the $(M,R)$ diagram in Fig.~\ref{MR} into three regimes (quantum, relativistic, classical). However, there are several other interesting lines in this diagram. For example, the vertical line 
$M=M_P$ is often assumed to mark the division between  elementary particles ($M <M_P$) and black holes ($M > M_P$), because one usually requires a black hole to be larger than its own Compton wavelength.

The horizontal line $R=R_P$ in Fig.~\ref{MR} is also significant because a simple heuristic argument suggests that quantum fluctuations in the metric should  become important below this. Since the energy density in a gravitational field with potential $\phi$ is $\rho_g \approx (\nabla \phi)^2/(8 \pi G)$, the mass associated with the gravitational energy within the volume $4 \pi R^3/3$ can only exceed  $\hbar c/R$ (as required by the HUP) for $\nabla \phi > \sqrt{ \hbar c G}/R^2$, so the metric fluctuations are 
 \begin{equation}
\Delta g/g \sim \phi/c^2 \sim R \nabla \phi/c^2 > R_P/R \, .
 \end{equation}
One therefore expects 
some form of spacetime foam for $R < R_P$ \cite{wheeler}. 
Quantum gravity effects should also be important whenever the density exceeds the Planck value,
 \begin{equation}
\rho_P = c^5/(G^2  \hbar) \sim 10^{94} \mathrm {g \, cm^{-3}} \, ,
 \end{equation}
corresponding to the sorts of  curvature singularities associated with the big bang or the centres of  black holes. This implies
 \begin{equation}
  R < (3M/ 4 \pi \rho_P) ^{1/3} \sim (M/M_P)^{1/3}R_P \, ,
  \label{planck3}
  \end{equation} 
which is well above the $R = R_P$ line in Fig.~\ref{MR} for $M \gg M_P$. So one might regard the combination of this line and  the $R=R_P$ line as specifying the boundary of  the ``quantum gravity" domain, as indicated by the shaded region in Fig.~\ref{MR}.  

Although we lack a detailed understanding of quantum gravity effects, the expressions for the quantum and black hole boundaries 
presumably break down as one appoaches their intersection point.  So although the Compton and Schwarzschild boundaries 
are represented as straight lines in the logarithmic plot of Fig.~\ref{MR}, this should only be regarded as an approximation. The proposal advocated in this paper 
requires that these boundaries must
join together smoothly in a way which depends upon their corrected forms. In the next two sections we discuss the possible modifications to the Compton and Schwarzschild expressions. 

\if
One unsatisfactory feature of Fig.~\ref{MR} is that it makes too clean a division between the quantum and relativistic domains. For example, 
it is well known quantum radiation is associated with the black hole an event horizon, so quantum effects are not really restricted to the region indicated by Eq.~\eqref{compton}. Hawking radiation is only important in practice (i.e. on a cosmological timescale) for $M < 10^{15}$g but that is still a long way from the Compton domain. This is another reason for suggesting that there may be some connection between the quantum and black hole boundaries. 
\fi

 \section{Generalized Uncertainty Principle}

Adler and colleagues~\cite{Adler} have discussed how the quantum boundary in Fig.~\ref{MR} might be modified as one approaches the Planck point from the left. They argue that the Uncertainty Principle should take the form
 \begin{equation}
 \Delta x > \hbar/ \Delta p + \alpha R_P^2 (\Delta p/ \hbar) 
\label{GUP1} 
\end{equation}
where $\alpha$ is a dimensionless constant which depends on the particular model and the factor of $2$ in the first term has been dropped. They describe this as the Generalized Uncertainty Principle (GUP) and offer a series of heuristic arguments for the second term in Eq.~\eqref{GUP1}. Subsequently, other authors have advocated this view \cite{chang} and considered its observational implications (eg. for the Lamb shift and Landau levels) \cite{das}.

\if
For example, the gravitational influence of the photon will accelerate the particle for a certain time, inducing a displacement of the form given by Eq.~\eqref{GUP1}. Alternatively, 
the spacetime distortion induced by the energy of the photon will produce spatial fluctuations of this form. Since one can also write the second term as $G \Delta p/c^3$, which is the Schwarzschild radius associated with an energy $c \Delta p$, one can also interpret this as meaning that this energy cannot be localized on a scale less than the associated black hole size. {\bf We return to this point later.}
Adler and colleagues~\cite{adler} have given several heuristic arguments for Eq.~\eqref{GUP1} and 
\fi

We first review some of these arguments, following the discussion in Ref.~\cite{adler1}.
The simplest one depends on a Newtonian analysis of gravitational effects and goes back to Ref.~\cite{mead}. A photon of frequency $\omega$ which comes within a distance $l$ of a particle will impart a gravitational acceleration $ a \sim G \hbar \omega /(c l)^2$ over a time $\Delta t \sim l/c$. It therefore induces a displacement
 \begin{equation}
 \Delta x_g \sim a (\Delta t)^2 \sim G \hbar \omega/ c^4 \sim G \Delta p/c^3 \sim R_P^2 \Delta p/ \hbar \, ,
 \label{Newton}
 \end{equation}
the uncertainty in the particle's momentum $\Delta p$ corresponding to the photon's momentum $\hbar \omega/c$. Assuming the uncertainties add linearly, the total $\Delta x$ therefore has the form indicated by Eq.~\eqref{GUP1}. So the imprecision in the position of the particle is due to the photon's momentum for low $\Delta p$ and its gravitational effect for high $\Delta p$.  The important point is that $\Delta x$ has a minimum at around the Planck scale.

\if
An equivalent argument is that the uncertainty in position due to the photon's gravitational potential at a distance of one wavelength is 
 \begin{equation}
\Delta x_g \approx \frac {\lambda \phi}{c^2} \approx \frac{G \hbar}{c^3 \lambda} \approx \frac{R_P^2}{\lambda} \, ,
 \end{equation}
so the total positional uncertainty is $\lambda + R_P^2/\lambda$.
This is equivalent to Eq.~\eqref{GUP1} if one identifies $\Delta p$ with $\hbar/ \lambda$. The gravitational field of the photon itself induces collapses for $\lambda < R_P$ or $\Delta p > c M_P$.
\fi

The reasoning leading to Eq.~\eqref{Newton} is Newtonian but Einstein's equations give an equivalent relativistic argument. A photon of momentum $p$ will produce a metric fluctuation $\Delta g_{\mu \nu}$ on a scale $R$ given by
 \begin{equation}
\frac { \Delta g_{\mu \nu}}{R^2} = \left(\frac {8\pi G}{c^4}\right)  \frac{p c}{R^3} \, .
\end{equation}
So if the probing photon is again assumed to impart an uncertainty $\Delta p \sim p$ in the momentum of the particle, the uncertainty in its position becomes
 \begin{equation}
 \Delta x_g \sim R  \Delta g_{\mu \nu} \sim 8 \pi G \Delta p/c^3 \sim R_P^2 (\Delta p/ \hbar) \, .
 \end{equation}
 A more precise calculation, differentiating between the transverse and longitudinal motions, gives the same result apart from numerical factors of order unity~\cite{Adler}. 

Variants of Eq.~\eqref{GUP1} can be found in particular theories -- for example, 
in non-commutative quantum mechanics or from general minimum length considerations \cite{maggiore}. The HUP is also modified in string theory because strings expand when probed at high energies, giving a GUP of the form
 \begin{equation}
 \Delta x > \hbar/ \Delta p + \alpha' (\Delta p/ \hbar) 
\label{GUPstring}
\end{equation}
where $ \alpha' \sim (10R_P)^2$ is the string tension \cite{veneziano}.
However, 
it is not clear that the second term can be related to black holes 
since the $M \gg M_P$ states are too stretched to undergo collapse. 
Indeed, in string theory a black hole is usually assumed to comprise {\it many} strings.

The GUP can also be derived in LQG because of polymer corrections in the structure of spacetime 
 \cite{ashtekar}. These have been studied by Hossain {\it et al.} \cite{hossain}, who find 
  \begin{equation}
 \Delta x > \frac {\hbar}{2 \Delta p} \left[1 - \frac {\lambda^2}{2} ( \Delta p)^2 + O(\lambda^4) \right] \, .
\label{hossain}
 \end{equation}
Here the factor of $2$ has been included and  $\lambda$ is a parameter of order $R_P/ \hbar$ associated with the polymer scale. The sign of the second term is negative in this model because the lattice structure reduces the uncertainty in position. However, this equation cannot apply for arbitrarily large $\Delta p$ since $\Delta x$ becomes negative for  $\Delta p > \sqrt{2}/ \lambda$. 
Note that Hossain {\it et al.} extract information
about the GUP from the LQG-inspired general polymeric quantization 
and their variables $\Delta x$ and $\Delta p$ are 
not the same as ours.
Their $x$ and $p$
correspond to the metric 
and the conjugate momentum, i.e. $g_{\mu \nu}$ and the extrinsic curvature $K_{\mu \nu}$, whereas 
ours correspond to
the position and momentum of a particle in spacetime. 

\if
The Hossain {\it et al.}  relation is also valid in our case because the metric is a solution of the polymeric Hamilton equations of motion 
inside the event horizon. In Ref.~\cite{extra}, 
where the solution was obtained for the first time, the phase space is parametrized by $c, p_c, b, p_b$.
[DEFINE] These variables are then replaced in LQG with the holonomies $h(c), p_c, h(b), p_b$. 
The new variables are the analog of  Eqs. (6) and (8a) in Hossain {\it et al.} but their relations (14, 15, 16) remain valid in our approach.]
\fi

\begin{figure}
 \begin{center}
 \includegraphics[height=4cm]{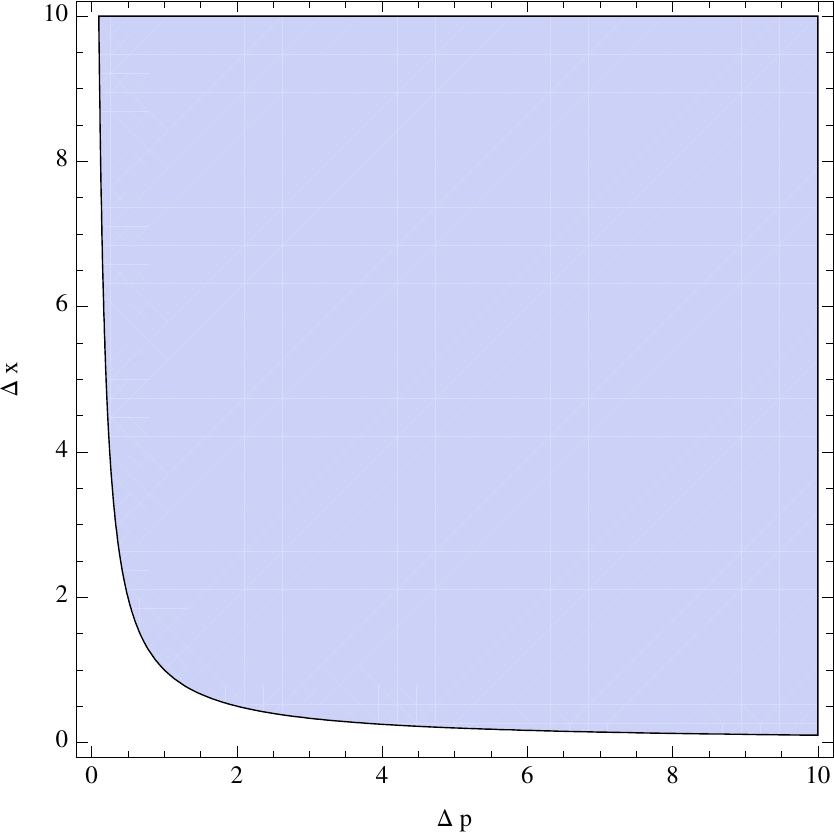}
  \includegraphics[height=4cm]{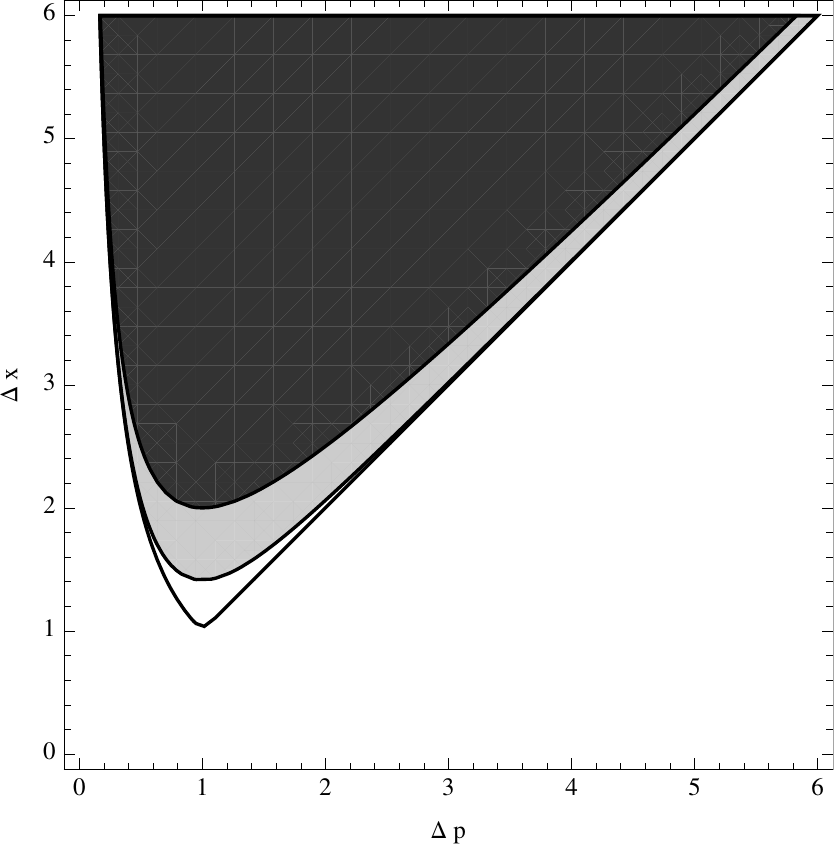}
   \end{center}
  \caption { \label{modesto1} This shows $\Delta x$ versus $\Delta p$ for the HUP (left)
 and the GUP (right) with $n$ =1, 2 and 20.}
  \end{figure}

Since the second term in Eq.~\eqref{GUP1} can be written as $\alpha G (\Delta p)/c^3$, it roughly  corresponds to the Schwarzschild radius for an object of mass $\Delta p/c$.  
Indeed, if we rewrite Eq.~\eqref{GUP1} using the substitution $\Delta x \rightarrow R$ and $\Delta p \rightarrow c M$, i.e. in the same way that one goes from Eq.~\eqref{UP} to the condition $R>R_C$, it becomes
 \begin{equation}
R > R_C' = \hbar/(Mc) + \alpha GM/c^2 \, .
\label{GUP2}
 \end{equation}
The expression on the right might be regarded as a generalized Compton wavelength and it asymptotes to the Schwarzschild form at large $M$, apart from a numerical factor.
However, it is unclear whether the advocates of the GUP intended it to be applied in this regime, so it would be more natural to write condition \eqref{GUP2} as
 \begin{equation}
R > R_C' = \frac{\hbar}{Mc} \left[ 1  + \alpha (M/M_P)^2 \right] \, .
\label{GUP3}
\end{equation}
The second term can be regarded as a correction as one approaches the Planck point from the left, this being small for $M \ll M_P$. On the other hand, the extension of this expression to all values of $M$ seems reasonable, since an outside observer cannot localize an object on a scale smaller than its Schwarzschild radius. This form of the GUP is shown by the $n=1$ curve in Fig.~\ref{modesto1}. 

\if
This corresponds to the condition
 \begin{equation}
\Delta x  > (2G/c^3) \Delta p \, ,
 \end{equation}
which leads to Eq.~\eqref{schwarzschild} if one uses the usual substitution $\Delta x \rightarrow R$ and $\Delta p \rightarrow c M$. 
\fi

A more general issue is that the usual Compton wavelength expression makes no sense for an object with $M \gg M_P$
since $\hbar /(Mc)$ is below the Planck scale. So one might wonder whether the GUP is relevant even for ordinary macroscopic objects. Of course, quantum effects are tiny in this regime, since one is well above the Compton line, but this is still an important issue of principle. 
One view might be that the Compton formula does not apply for {\it composite} objects but that raises the issue of what is meant by ``composite''. For example, the Compton wavelength of a proton is usually taken to be $ \hbar/(m_p c) \sim 10^{-13}$cm, even though the proton is made of three quarks with much larger individual Compton wavelengths. And should a black hole be regarded as composite or discrete?
 At the very least, one should worry about the application of  the HUP for $M \gg M_P$. 
 
In a sense, this concern underlies the whole issue of quantum decoherence. For example, Kay's modification of non-relativistic quantum mechanics~\cite{kay} leads to a GUP for an object of mass $M$ and radius $R$ of the form
 \begin{equation}
 \Delta p \ge { \hbar \over 2 \Delta x} +  9 \left( {M \over M_P} \right)^2 {\hbar \Delta x  \over R^2}  \, ,
\end{equation}
which gives a lower limit on $\Delta p$ rather than $\Delta x$. 
This result is only applicable in the Newtonian limit and for wave
functions whose spread is smaller than $R$ but if one makes the substitution $\Delta x \rightarrow R$ and $\Delta p \rightarrow c M$, one obtains the relation
\begin{equation}
 R \ge {\hbar \over  2 Mc}  [1+ 18 (M/M_P)^2]  \, .
\end{equation}
This looks like the Compton line for $M << M_P$ but tends to roughly the black hole line for $M >> M_P$, which is similar to our proposal.

These considerations  
suggest that  there is a different kind of positional uncertainty for objects larger than the Planck mass, with $\hbar/ \Delta p$ being replaced by the (much larger) Schwarzschild radius for the mass $M = \Delta p/c$. Postulating that no external observer  can localize an object more precisely than the event horizon might also be interpreted in terms of the 
time distortions near a black hole, since the event horizon only forms asymptotically from the perspective of external observer. One might further speculate that there is a deeper level of uncertainty below the Planck density line in Fig.~\ref{MR}, 
since the notion of space as a continuum breaks down here. 

Note that there is an important distinction between the uncertainty in the  position of an object which falls inside a black hole and the position of the black hole itself. For example, dynamical observations might be able to determine the position of a black hole's {\it centre of mass} more precisely than its radius. This distinction will also arise when we consider quantum emission by black holes.  


Various caveats should be stressed at this point. First, Eq.~\eqref{GUP2} only corresponds {\it exactly} to the Schwarzschild condition for $M \gg M_P$ if $\alpha =2$ and there is no obvious reason for this in the heuristic derivations of the GUP given above. Second, 
\if
despite the form of Eq.~\eqref{GUPstring}, the link with black holes would not seem to pertain in string theory because the strings are too elongated to form black holes.
Another concern is that 
\fi
Eq.~\eqref{GUP1} assumes that the two uncertainties add {\it linearly}. It is possible that the operator structure of quantum theory requires this. On the other hand, 
since they are independent, it might be more natural to assume that they add quadratically:
 \begin{equation}
 \Delta x > \sqrt {(\hbar/ \Delta p)^2 + (\alpha R_P^2 \Delta p/ \hbar)^2} \, .
 \end{equation}
This would correspond to a generalized Compton wavelength
 \begin{equation}
R_C' = \sqrt{ (\hbar /Mc)^2  + (\alpha GM/c^2)^2 } \, ,
\label{quad}
\end{equation}
which is different from the expression given by Eq.~\eqref{GUP2}. The approximation \eqref{GUP3} for $M \ll M_P$ is then replaced with
\begin{equation}
R_C' \approx \frac{\hbar}{Mc} \left[ 1  + \frac{\alpha^2}{2} (M/M_P)^4 \right] \, ,
\label{quadapprox}
\end{equation} 
implying that the deviation from the HUP falls off fa ster than before
 as $M$ falls below $M_P$. While the heuristic arguments given above indicate the form of the two uncertainty terms, they do not specify how one combines them.

In fact, depending on one's quantum gravity model, there could be many possible curves which smoothly interpolate between the Compton and Schwarzschild limits. For example, one might envisage a GUP of the form
 \begin{equation}
 \Delta x >  [(\hbar/ \Delta p)^n + (\alpha R_P^2 \Delta p/ \hbar)^n] ^{1/n}
 \label{newGUP}
 \end{equation}
for some integer $n$, corresponding to the condition
 \begin{equation}
 R > R_C' = [(M_P/M)^n + (\alpha M/M_P)^n]^{1/n} R_P \, .
\label{GUPn}
 \end{equation}
This gives the approximation
\begin{equation}
R > R_C' \approx  \frac{\hbar}{Mc} \left[ 1  + \frac{\alpha^n}{n} (M/M_P)^{2n} \right] 
\end{equation} 
for $M \ll M_P$.
\if
 or, equivalently, 
\begin{equation}
 \Delta x > \frac{\hbar}{ \Delta p} + \frac{\alpha^n}{n} R_P^{2n} \left( \frac{\Delta p}{\hbar} \right)^{2n-1} 
\end{equation}
for $\Delta p \ll c M_P$.
\fi
The version of the GUP given by Eq.~\eqref{GUP3} then corresponds to $n=1$, 
whereas Eq.~\eqref{quadapprox} corresponds to $n=2$, but 
in principle any value of $n$ would be possible.
Each expression would correspond to a different point in Fig.~\ref{modesto1} where $R_C'$ bottoms out, the minimum occurring at $M = \alpha^{-1/2}M_P$ and $R = 2^{1/n} \alpha^{1/2} R_P$.  The form of the GUP is indicated for three values of $n$.

More generally, one might consider any function $R_C'(M)$ which scales as $M$ for $M \gg M_P$ and $M^{-1}$ for $M \ll M_P$. These would all have the required asymptotic forms 
but differ near the Planck scale itself. A Taylor expansion  for $\Delta p \ll c M_P$ might then give a GUP of the form
\begin{equation}
 \Delta x > \hbar/ \Delta p + \sum_{n} k_n R_P^{n+1} (\Delta p/ \hbar)^n 
\label{taylor} 
\end{equation}
for dimensionless constants $k_n$. Measurements of these constants might then provide an experimental probe of the GUP. One factor discriminating between different forms of $R_C'(M)$ might be the requirement of Lorentz-invariance. 

\section{Generalized Event Horizon}

Discussions of the GUP usually focus on what happens to the left of the Planck point in Fig.~\ref{MR}. However, the GUP also has important implications for the black hole horizon size, as can be seen by  examining what happens as one approaches the intersect point from the right. In this limit, 
it is natural to write Eq.~\eqref{GUP2} as
 \begin{equation}
R > R_S' = \frac{\alpha GM}{c^2} \left[ 1  + \frac{1}{\alpha} (M_P/ M)^2 \right] \, ,
\end{equation}
which represents a small perturbation to the Schwarzschild radius for $M \gg M_P$ if one puts $\alpha =2$. None of the heuristic arguments require $\alpha =2$, so one has to assume this at the outset. Alternatively, one might envisage a somewhat different one-parameter form:
 \begin{equation}
R > R_S' = \frac{2 GM}{c^2} \left[ 1  + \beta (M_P/ M)^2 \right] 
\label{GUP4}
\end{equation}
for some constant $\beta$. One might then regard this expression as defining a Generalized Event Horizon (GEH). 

Since the GUP provides an extension of the quantum boundary into the relativistic domain, 
this raises the question of whether there is an equivalent extension of the relativistic boundary into the quantum domain. This is a natural inference since  Eq.~\eqref{GUP4} asymptotes to the Compton scale at {\it small} $M$ in the same way that the GUP bound tends to the Schwarzschild scale at {\it large} $M$. 
At first sight it might seem unphysical to contemplate a black hole with $M < M_P$ since it is smaller 
than its own Compton wavelength.
However, as described below, this problem is resolved in a rather surprising way in LQG.

The GEH concept also arises in more general forms of the GUP. In particular, if the uncertainties add quadratically, one can regard Eq.~\eqref{quad} as defining $R_S'$ as well as $R_C'$ and this leads to the approximation 
 \begin{equation}
R _S' \approx \frac{\alpha GM}{c^2} \left[ 1  + \frac{1}{2 \alpha^2} (M_P/ M)^4 \right] 
\end{equation}
for $M \gg M_P$. As in the linear case, one might then consider the alternative form
 \begin{equation}
R _S' \approx \frac{2 GM}{c^2} \left[ 1  + \beta(M_P/ M)^4 \right]  
\label{GUP5}
\end{equation}
for some parameter $\beta$. Recently it has been discovered that a model inspired by LQG permits the existence of a new type of black hole whose horizon size has precisely the form \eqref{quad}. 
The model is discussed in more detail in Sect.~\ref{metric}
but the crucial point
is that the physical radial coordinate $R$ is not the Schwarzschild coordinate $r$ but related to it by 
\begin{equation}
R = \sqrt{ r^2 + \frac {\xi^2 R_P^4}{r^2} } \, ,
\label{newR}
\end{equation}
where $\xi$ is a constant of order unity.
$R$ is {\it physical} in the sense that the area of the 2-sphere with constant $R$ is $4 \pi R^2$. 
The two coordinates are almost the same at large $r$ 
but radically different at small $r$.
Indeed, while $r$ spans the range  $0$ to $\infty$, $R$ has a minimum of $\sqrt{2 \xi} R_P$ at $r= \sqrt{\xi} R_P$ and goes to infinity in both limits. 
The central singularity of the Schwarzschild solution is therefore replaced with another asymptotic region, so the collapsing matter bounces and the black hole becomes part of a wormhole.

If one puts $r=2GM/c^2$ in Eq.~\eqref{newR}, corresponding to the position of the event horizon, one obtains
\begin{equation}
R_S' = \sqrt{ (2GM/c^2)^2  + (\xi \hbar /2Mc)^2 } \, .
\label{quad2}
\end{equation}
Providing $\xi =2$, this equation
has the implication that the black hole size {becomes the Compton wavelength for $M \ll M_P$.
There is no {\it a priori}
reason for taking $\xi = 2$, so this is analogous to the assumption that $\alpha =2$ in the GUP case. Equation~\eqref{newR} therefore has three important cosequences: (1) it removes the singularity; (2) it permits the existence of black holes with $M \ll M_P$; and (3) it allows a unified expression for the Compton and Schwarzschild scales. Indeed, it seems remarkable that the purely geometrical condition \eqref{newR} implies the quadratic version of the GUP given by Eq.~\eqref{quad}. 

The above discussion suggests that one should try to {\it generalize} and {\it unify} the standard expressions for $R_C$ and $R_S$ by seeking a function of $M$ which asymptotes to the Compton wavelength for $M \ll M_P$ and the Schwarzschild radius for $M\gg M_P$:
 \begin{equation}
R_C' = R_S'  \approx 
\begin{cases}
\hbar/(Mc) & (M \ll M_P) \\
2GM/c^2 & (M \gg M_P) \, .
\end{cases}
\label{GEH}
\end{equation}
This proposal is termed the 
Black Hole Uncertainty Principle 
(BHUP) correspondence, although it is important to stress that one only obtains the {\it exact} Compton and Schwarzschild expressions asymptotically for particular values of
$\alpha$ and $\beta$.

We postulate that a similar result may apply in other theories of quantum gravity
\if
For example, a formula like Eq.~\eqref{newR} might also arise in  
string theory because the energy of the string has two components: one corresponds to the string excitation mode and scales as $r^{-1}$; the other corresponds to  
the string winding mode and scales as $r$.  However, these energies should add linearly, so one would not obtain the quadratic expression  corresponding to Eq.~\eqref{newR}. Also, while Eq.~\eqref{GUPstring} formally resembles Eq.~\eqref{GUP1}, the second term cannot correspond to a black hole because  the string is too elongated, 
In a more general theory, therefore, 
\fi
if one replaces Eq.~\eqref{quad2} with any expression which reduces to the Compton and Schwarzschild forms asymptotically. For example, by analogy with Eq.~\eqref{GUPn}, one might consider the power-law expression
\begin{equation}
R_S' = [ (2GM/c^2)^m  + (\beta \hbar /Mc)^m ]^{1/m} 
\label{quad3}
\end{equation}
with some integer $m$, leading to the approximation
 \begin{equation}
R _S' \approx \frac{2GM}{c^2} \left[ 1  + \frac{\beta^m}{m} (M_P/ M)^{2m} \right] 
\end{equation}
for $M \gg M_P$. 
\if
In this case, Eq.~\eqref{newGUP} would imply
\begin{equation}
\Delta x \sim (M^m + M^{-m})^{1/m} \sim [ 1/ (\Delta p)^n + (\Delta p)^n]^{1/n} \, .
\label{quad4}
\end{equation}
\fi
This is only compatible with the equivalent generalization of the GUP, given by Eq.~\eqref{newGUP}, and
the identification $\Delta p \rightarrow Mc$ for all ranges of $M$ 
if $m=n$. 
In particular, LQG corresponds to the special case $m=n=2$. 

Whatever the form of $R_S'(M)$, the important qualitative point is that the BHUP correspondence requires that one associates $R_S'$ with a positional uncertainty $\Delta x$. Therefore $R_S'$ must have the same functional dependence on $M$ as $\Delta x$ has on $\Delta p$. 
By analogy to Eq.~\eqref{taylor}, one might then expect the GEH for $M \gg M_P$ to have the Taylor expansion
\begin{equation}
R_S' > 2GM/c^2  + R_P  \sum_{n} k_n (M/M_P)^{-n} 
\end{equation}
for constants $k_n$. The important point is that the transition is {\it smooth} and this is what provides the link between the Uncertainty Principle and black holes. 
One might regard the nature of this transition as a fundamental signature of any theory of quantum gravity. 

\if
It should be stressed that we are not claiming that {\it all} objects on the Compton line 
are black holes -- just that some kinds of black hole reside on this boundary. The relationship between black holes and elementary particles in LQG is discussed in Sec.~\ref{metric}.

\fi

\if
Note that in terms of the coordinate $R$, there is another Planck density line, given by [NO!]
  \begin{equation}
  R \approx (m/M_P)^{-1/3}R_P \, ,
  \end{equation} 
which is the reflection of the Planck density line in Fig.~\ref{MR}. This extends the striking symmetry between the macroscopic and microscopic domains.
\fi

\if
Note that the quantity $R$ is a measure of the proper distance in the transverse direction. In the radial direction, the differential relationship between the proper distance $\tilde R$ and $r$ may be more complicated but one can argue on general grounds that both must have the asymptotic differential form
\begin{equation}
\frac{\Delta \tilde R}{\Delta r} \approx \frac{\Delta R}{\Delta r} \approx
\begin{cases}
1 & (r \gg r_P) \\
(r/R_P)^{-2} & (r \ll r_P) \, .
\end{cases}
\label{radial}
\end{equation}
In particular, this result will be shown to apply in LGQ in Sec.~\ref{metric}. However, there are some ranges of $r$ in which the radial and transverse proper distances have a different dependence on $r$, so the uncertainty in position $\Delta x$ is in principle direction-dependent.
\fi

\if
One consequence of Eq.~\eqref{newR} is that the Planck density condition becomes 
\begin{equation}
M = \left( \frac{4 \pi  \rho_P }{3} \right) \left( r^2 + \frac{R_P^4}{r^2} \right)^{3/2} \, .
\end{equation}
This leads to a quadratic equation in $r^2$,
\begin{equation}
r^4 - \left( \frac{3M}{4 \pi \rho_P} \right)^{2/3} r^2 + R_P^4 = 0 \, .
\end{equation}
One root is 
\begin{equation}
r \approx \left( \frac{3M}{4 \rho_P} \right)^{1/3}  \approx R_P \left( \frac  {M} {M_P}\right)^{1/3} \; (M > M_P) \, ,
\end{equation}
which corresponds to Eq.~\eqref{planck}. The other root is
\begin{equation}
r \approx R_P^2 \left( \frac {4 \pi \rho_P} {3M} \right)^{1/3} \approx R_P \left( \frac  {M} {M_P}\right)^{-1/3}  \; (M < M_P)\, ,
\label{planck2}
\end{equation}
which appears in Fig.~\ref{MR} as the reflection of the first Planck density line about $M=M_P$. The significance of this second root will become clear later. The exact solution of the quadratic equation gives a smooth interpolation between these two regimes. [MUST CORRECT ABOVE DISCUSSION SINCE VOLUME IS $R \tilde R^2$ RATHER THAN $\tilde R^3$. ALSO DEPENDENCE OF $\Delta x$ ON $M$ IS DIFFERENT IN RADIAL AND TRANSVERSE DIRECTIONS!]
\fi

\section {GUP and black hole thermodynamics}

A particularly interesting application of the GUP concerns the link with black hole thermodynamics. 
Although Hawking's discovery of black hole radiation provides fundamental insights into quantum gravity, it is at first sight rather surprising since the black hole region in Fig.~\ref{MR} is well above
the Compton line. Quantum radiance may be small for macroscopic black holes but it is still important in principle and even black holes with 
$M \sim 10^{15}$g, which evaporate at the present epoch, are  far from the Compton line. However, a perspective in which the black hole boundary {\it becomes} the quantum boundary in the macroscopic domain accommodates Hawking's result very naturally.

Let us first recall the link between black hole radiation and the HUP.  This arises because we can obtain the black hole temperature for $M \gg M_P$ by 
identifying $\Delta x$ with the Schwarzschild radius and $\Delta p$ with some multiple of the black hole temperature:
\begin{eqnarray}
kT = \eta c \Delta p = \frac{ \eta \hbar  c}{\Delta x} = \frac{\eta \hbar c^3}{ 2 G M} \, .
\label{temp}
\end{eqnarray}
This gives the precise Hawking temperature 
if we take $\eta = 1/(4\pi)$. Since 
$T \sim M_P^2/M$, the mass associated with this temperature is roughly the ``dual'' of the black hole mass, in the sense that these masses are reflections about the line $M=M_P$ in Fig.~\ref{MR}.  
It should be stressed that the application of the HUP in this argument involves both the particle emitted by the black hole and the black hole itself. The second equality in Eq.~\eqref{temp} relates to the emitted particle and assumes that $\Delta x$ and $\Delta p$ satisfy the HUP. The third equality relates to the black hole and assumes that $\Delta x$ is the Schwarzschild radius. 

Both the above assumptions require $M \gg M_P$ but even
in this regime the GUP requires a correction to the standard temperature formula. 
If we adopt the GUP in the most general form \eqref{newGUP} and associate $\Delta p$ with $T$ and $\Delta x$ with $R_S$, as before, then we obtain 
\begin{equation}
T = {M c^2 \over 2^{1+1/n} \pi  \alpha k} \left(1- \sqrt{1- 2^{2-2n} \alpha^n \left( {M_P \over M} \right)^{2n} } \right)^{1/n} \, .
\label{adlertemp}
\end{equation}
This can be approximated by 
\begin{equation}
T  \approx {\hbar c^3 \over 8\pi G k M} \left[ 1 + \frac{ \alpha^n}{n 2^{2n}} \left( {M_P \over M} \right)^{2n} \right] 
\end{equation}
for $M \gg M_P$, with the last term representing a small perturbation to the Hawking prediction.
However, the assumption $\Delta x \approx R_S $ also breaks down as $M$ falls towards $M_P$ and this entails a correction of the same order as the GUP itself. Indeed, Eq.~\eqref{quad3} implies that the term $\alpha ^n$ in the above analysis must be replaced by $\alpha^n - (2 \beta)^n$. 

While the exact modification is complicated and model-dependent in the $M \gg M_P$ regime, 
we can easily understand
the temperature behavior for $M \ll M_P$, a possibility which at least arises in LQG.
In this case, 
Eq.~\eqref{GEH} implies $\Delta x \approx \hbar/ (Mc)$ and so Eq.~\eqref{temp} becomes
\begin{eqnarray}
kT = \eta c \Delta p = \frac{ \eta \hbar  c}{\Delta x} \approx \eta M c^2 \, .
\label{subtemp}
\end{eqnarray}
Since this is less than the Planck temperature, the second equality still applies to a good approximation, as required for consistency. However, one can use another argument which gives a different result. 
Since the temperature is determined by the black hole's surface gravity \cite{hawking}, Eq.~\eqref{GEH} suggests 
\begin{eqnarray}
T \propto \frac{GM}{R_S'^2} \propto 
\begin{cases}
M^{-1} & (M \gg M_P) \\
M^3 & (M \ll M_P) \, ,
\end{cases}
\label{GUPtemp}
\end{eqnarray}
so the black hole temperature should scale as $M^3$ rather than $M$ for $M \ll M_P$.
Since we cannot be certain that the temperature is identified with the surface gravity in this regime, this raises the issue of whether we believe Eq.~\eqref{subtemp} or \eqref{GUPtemp} for $M \ll M_P$. Both equations predict that the temperature of a black hole deviates from the Hawking expression when its mass falls below $M_P$ and that it never goes above the Planck scale. 
But which prediction is correct?

Obviously one cannot answer this question definitively without having a proper theory of quantum gravity. However, one heuristic reason for believing Eq.~\eqref{GUPtemp} is that it implies that the entropy associated with a black hole is
 \begin{equation}
S \propto \int \frac {dM}{T} \propto 
\begin{cases}
M^2 & (M \gg M_P) \\
M^{-2} & (M \ll M_P) \, .
\end{cases}
 \end{equation}
So the entropy scales as the area in both asymptotic limits, although it is negative for $M \ll M_P$. If $T \propto M$ for $M \ll M_P$, $S$ has a logarithmic dependence on $M$ in this regime and this suggests that Eq.~\eqref{subtemp} is less plausible. We therefore favour the mass dependence implied by Eq.~\eqref{GUPtemp} and this is indicated in Fig.~\ref{logtemp}. Note that the GUP changes the relationship between the entropy and area of a black hole somewhat anyway, even for $M \gg M_P$, the entropy typically acquiring an extra logarithmic term~\cite{camellia}.

The apparent inconsistency between Eq.~\eqref{subtemp} and Eq.~\eqref{GUPtemp} for $M \ll M_P$ may arise because
one needs to distinguish between the emitted particle (em) and the black hole itself (BH). 
The quantity $\Delta p$ appearing in Eqs.~\eqref{temp} and \eqref{subtemp} always refers to $(\Delta p)_{em}$ and this is 
different from $(\Delta p)_{BH}$ since the latter scales as $M$.
Equation~\eqref{GUPtemp} thus implies
\begin{equation}
\frac {(\Delta p)_{em} }{ (\Delta p)_{BH} }  \approx
\begin{cases}
(M/M_P)^{-2} & (M \gg M_P) \\
(M/M_P)^{2} & (M \ll M_P) \, ,
\end{cases}
 \end{equation}
although this relationship is not required in the derivation of $T$. The relationship between $(\Delta x)_{em}$ and $(\Delta x)_{BH}$, which {\it is} required, is less clearcut. However, since these scale as $T^{-1}$ and $R_S'$, respectively, Eqs.~\eqref{GEH} and \eqref{GUPtemp} suggest
 \begin{equation}
\frac {(\Delta x)_{em} }{ (\Delta x)_{BH} }  \approx
\begin{cases}
1 & (M \gg M_P) \\
(M/M_P)^{-2} & (M \ll M_P) \, .
\end{cases}
\label{strange}
 \end{equation}
\if
More precisely, the mass and length scale of the emitted particle are given by 
\begin{equation}
M_{em} \sim ( \Delta p)_{em} \propto M^{-1} , \quad R_{em} \sim (\Delta x)_{em} \propto M
\end{equation}
for $M \gg M_P$ and by
\begin{equation}
M_{em} \sim (\Delta p)_{em} \propto  M^{\zeta}, \quad R_{em} \sim (\Delta x)_{em} \propto M^{-\zeta} 
\end{equation}
for $M \ll M_P$, where $\zeta$ is $1$ according to Eq.~\eqref{temp} and $3$ according to Eq.~\eqref{GUPtemp} . In both cases, one has $R_{em} \propto M_{em}^{-1}$
since the energy of the emitted particle is sub-Planckian. On the other hand, the GUP implies that the black hole itself has
\begin{equation}
(\Delta p)_{BH} \propto M, \quad  (\Delta x)_{BH} \propto M
\end{equation}
for $M \gg M_P$ and 
\begin{equation}
 (\Delta p)_{BH} \propto M, \quad (\Delta x)_{BH} \propto M^{-1}
\end{equation}
for $M \ll M_P$.

Since $\Delta x_{em}$ is in some sense radial, Eq.~\eqref{radial} suggests
 \begin{equation}
\Delta x_{em} \sim r^{-2} \Delta x_{BH}  \sim M^{-3}
 \end{equation}
where the last step uses $r  \sim M$ and $\Delta x_{BH} \sim M^{-1}$, corresponding to the size of the black hole. By this argument, Eq.~\eqref{temp} naturally implies Eq.~\eqref{GUPtemp}.
[THIS CALCULATION MAY BE A FUDGE BUT IT GETS CORRECT ANSWER!] 

Consistency requires 
 \begin{equation}
 (\Delta x)_{em} = ( \Delta x)_{BH}
  \end{equation} 
 and 
 \begin{equation}
\frac {(\Delta p)_{em} }{ (\Delta p)_{BH} }  \approx
\begin{cases}
(M/M_P)^{-2}  & (M \gg M_P) \\
1 & (M \ll M_P) 
\end{cases}
 \end{equation}
if $\zeta =1$ but
 \begin{equation}
\frac {(\Delta x)_{em} }{ (\Delta x)_{BH} }  \approx
\begin{cases}
1 & (M \gg M_P) \\
(M/M_P)^{-2} & (M \ll M_P) 
\end{cases}
\label{strange}
 \end{equation}
and
 \begin{equation}
\frac {(\Delta p)_{em} }{ (\Delta p)_{BH} }  \approx
\begin{cases}
(M/M_P)^{-2}  & (M \gg M_P) \\
(M/M_P)^2 & (M \ll M_P) 
\end{cases}
 \end{equation}
if $\zeta =3$. 
\fi
Note that Eq.~\eqref{newR} implies the differential relation
\begin{equation}
\frac{\Delta R}{\Delta r} \approx
\begin{cases}
1 & (r \gg r_P) \\
(r/R_P)^{-2} & (r \ll r_P) 
\end{cases}
\label{radial}
\end{equation}
and this resembles Eq.~\eqref{strange} if one puts $r=2GM/c^2$. 
This suggests that $(\Delta x)_{BH}$ and $(\Delta x)_{em}$ should be identified with 
$\Delta r$ and 
$\Delta R$, respectively. 
One justification for this might be that it is natural to relate $(\Delta x)_{BH}$ with $r$ because it refers to observations in our asymptotic space, while $(\Delta x)_{em}$ is associated with $R$ because the black hole emission is measured in the other asymptotic space.

\if
Another possible justification for 
$(\Delta x)_{BH}\sim R_S$ and $(\Delta x)_{em}\sim R_P$ when $M<<M_P$ 
is as follows. The matter located within the black hole is inside the horizon and thus can only be located within a precision of the the size of the horizon $R_S$. 
However, the radiated particle which is 
observed at infinity must come through the Planck-sized wormhole throat behind which the horizon is hidden, so the uncertainty in its position is of the order of the Plank length.
\fi

\begin{figure}
 \begin{center}
 \includegraphics[height=4cm]{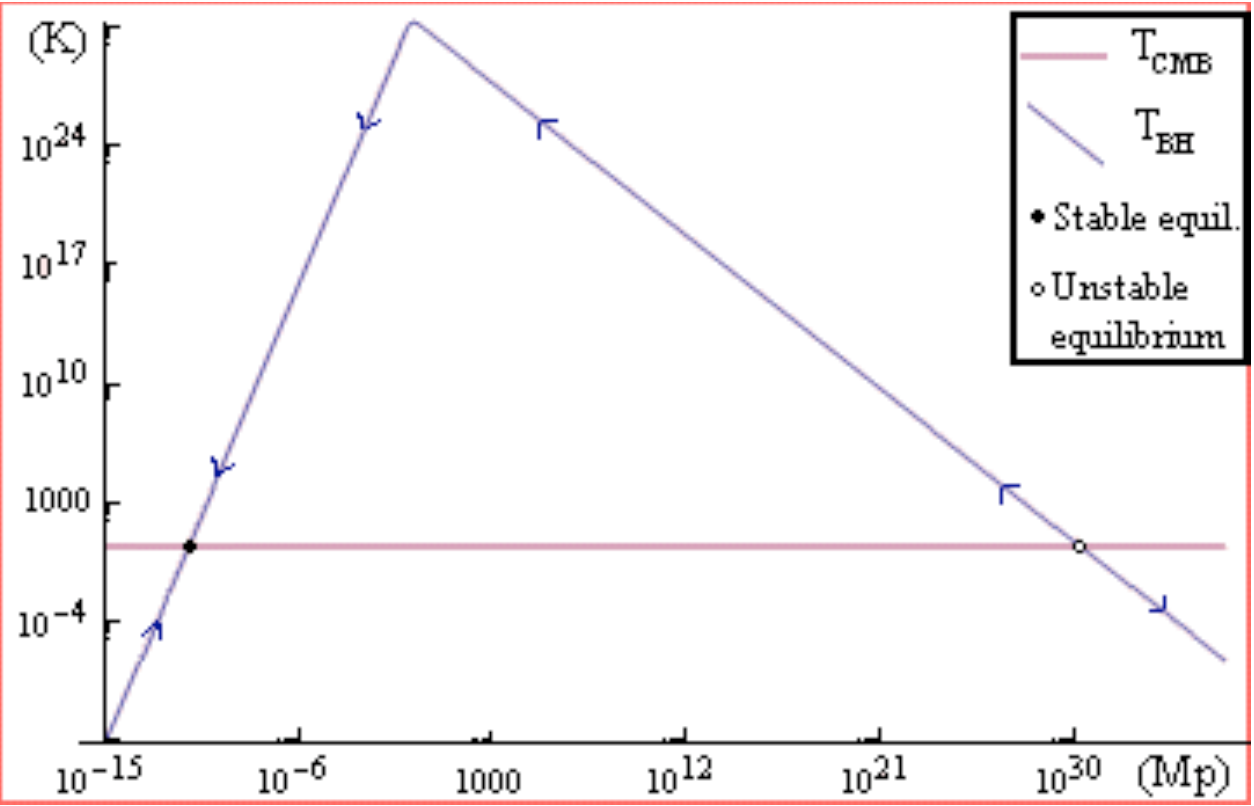}
   \end{center}
\caption{Showing the dependence of temperature on the black hole mass predicted by the GUP and comparing this to the CMB temperature.} 
\label{logtemp}
\end{figure}

In any case, there is certainly an important difference between sub-Planckian and super-Planckian black holes. This is because a collapsing body with $M \gg M_P$ reaches the Planck density before it collapses within a Planck volume, whereas one with $M \ll M_P$ falls within a Planck volume first and is therefore hidden behind a wormhole throat of diameter $\sim R_P$. 
Since any particle which is 
radiated into our Universe must also come through this throat, this
suggests that the spectrum of particles detected in the two asymptotic regions should be different. 
Replacing  $R_S'$ with $R_P$ in Eq.~\eqref{GUPtemp} implies that the temperature associated with the wormhole throat itself is $T \propto M$, which corresponds to the prediction of Eq.~\eqref{subtemp}.

The above discussion 
might be contrasted with the argument of Adler {\it et al.} \cite{Adler}.
By associating $\Delta p$ and $\Delta x$ with the temperature and Schwarzschild radius, respectively, they derive the  temperature 
\begin{equation}
T = {M c^2 \over 4\pi k} \left(1- \sqrt{1- \frac{\alpha M_P^2}{M^2}} \right) \, .
\label{adlertemp2}
\end{equation}
This corresponds to Eq.~\eqref{adlertemp} with $n=1$ but they apply it for {\it all} values of $M$.
\if
to the standard Hawking form 
\begin{equation}
T_{BH} = {c^2 M_P^2 \over 8 \pi k M} =  {\hbar c^3 \over 8\pi G k M}
\label{hawking}
\end{equation}
for $M \gg M_P$ but it remains finite instead of diverging at the Planck mass itself. 
Specifically, putting $\Delta p \sim T$ and $ \Delta x \sim M$ into Eq.~\eqref{GUP1} yields a quadratic equation in $T$ with solutions
 \begin{equation}
T \approx Mc^2 [1 - \sqrt {1-M_P^2/M^2} ] \, . 
\label{adlertemp}
 \end{equation}
We have seen that this is close to the standard form for $M \gg M_P$ but 
\fi
Their expression peaks with $T \sim M_P$ at $M \sim M_P$ and then becomes complex for $M < M_P$.  They
infer that evaporation ceases at the Planck mass, leading to stable relicts. 
The discrepancy 
arises because they assume $\Delta x \propto M^{-1}$ for $M \ll M_P$, whereas we assume $\Delta x \propto M$. The temperature never goes complex in our scenario but there are still {\it effectively} stable relics since the evaporation timescale becomes longer than the age of the universe for sufficiently small $M$ \cite{Modesto:2008im}. Just as the HUP stabilizes the ground state of the hydrogen atom, so the GUP stabilizes the ground state of a black hole.
This is also reminiscent of the suggestion that the black hole surface area has a discrete spectrum, with a uniform spacing determined by the area-entropy relationship \cite{hod}. 
However, it is not clear how this relates to the present proposal.

\if
In fact, the exact analysis would give
\begin{equation}
T =  \sqrt{M^2 + M_P^4/M^2} \pm \sqrt {M^2 + M_P^4/M^2 - 4 M_P^2} \, . 
\label{exacttemp}
 \end{equation}
This implies $T \propto M^{-1}$ for $M \gg M_P$ and $T \propto M$ for $M \ll M_P$, although the latter becomes $T \propto M^{3}$ if one invokes Eq.~\eqref{radial}.

The relationship $\Delta p \propto M^{-1}$ refers to the former, whereas the relationship $\Delta p \propto M$ refers to the latter.
Since Eq.~\eqref{logtemp} implies that the black hole temperature never goes  
above the Planck value, even for $M \ll M_P$, the emitted particles are always sub-Planckian. This means that one always has $T \sim \Delta p \sim 1/\Delta x$, which justifies the second equality in   
Eq.~\eqref{temp}. However, the mass of the black hole may be either sub-Planckian or super-Planckian and the third equality in   
Eq.~\eqref{temp} only applies in the latter case. In the former case, one would infer $\Delta x \propto M^{-1}$ and hence $T \propto M$. Only the black hole itself, not the emitted particles, can be in the  regime for which $\Delta p \sim \Delta x$. 

The total number of particles emitted is
\begin{eqnarray}
N_{em} \approx \left( \frac{Mc^2}{kT} \right)  \sim 
\begin{cases}
(M/M_P)^2 & (M \gg M_P) \\
(M/M_P)^{-2} & (M \gg M_P) \, .
\end{cases}
\end{eqnarray}
The relationship $(\Delta x)_{BH} \sim N_{em} (\Delta x)_{em}$ would then be compatible with the form of the GUP required by Eq.~\eqref{GUP1}.  [REALLY?] 
\fi


 \section{Loop Black Holes}
\label{metric}

One approach to quantum gravity, Loop Quantum Gravity
(LQG) \cite{LQGgeneral},
has given rise to
models which describe the very early universe. This simplified
framework, which uses a mini-superspace approximation, has been shown
to resolve the initial singularity problem \cite{Bojowald}.
 A black hole metric in this model, known as the loop black hole (LBH)
 \cite{Modesto:2008im}, has  a self-duality property
that removes the singularity and replaces it with
another asymptotically flat region. The
thermodynamic properties of these self-dual black holes have been
examined in \cite{
poly} and
 the dynamical aspects of the collapse and
evaporation were studied in \cite{Hossenfelder:2009fc}. 
The black hole spacetime has also been studied in a mini-superspace  \cite{brannlund}
and midi-superspace \cite{GP}  reduction of LQG.

LQG is based on a canonical quantization of the Einstein equations written in terms of the Ashtekar variables \cite{AA}, i.e. in terms of an $SU(2)$ 3-dimensional connection $A$ and a triad $E$. The basis states of LQG are then closed graphs whose edges and vertices are labeled by irreducible $SU(2)$ representations and $SU(2)$ intertwiners, respectively. (See \cite{LQGgeneral}
for a review.) The
edges of the graph represent quanta of area $8 \pi \gamma R_P^2 \sqrt{j(j+1)}$, where $j$ is a half-integer $SU(2)$ representation of the edge label
and $\gamma$ is a parameter of order one, called the Immirzi parameter. The vertices of the graph represent quanta of $3$-volume. One important consequence of this is 
that the area is quantized, with the smallest possible 
area  being
\begin{equation}
A_{\mathrm{min}} = 4 \pi \sqrt{3} \gamma R_P^2 \, .
\end{equation}
However, one should not take this exact value too seriously for various reasons: 
(1) the value of $\gamma$ is not definite and the consensus on its value has change a few times already; (2) other Casimirs are possible besides $\sqrt{j(j+1)}$;  (3) we are looking for a minimum area for a closed surface but the spin-network is probably a closed graph, so it is likely that at least two edges cross the surface, in which case the minimum area is doubled;
(4) if we consider a surface enclosing a non-zero volume, LQG stipulates that at least one 4-valent vertex must be present, in which case we might have four edges intersecting the surface, 
quadrupling the minimum area.
 
 We will parametrize our ignorance 
 with a parameter $\beta$ defined by
\begin{equation}
A_{\rm min} = 
4\pi\gamma\beta \sqrt{3} \, R_P^2\approx 20 \gamma \beta R_P^2 \,.
\end{equation}
We then introduce another parameter
\begin{equation}
a_o=A_{\rm min}/8\pi= \sqrt{3}\, \gamma\beta R_P^2/2 \, .
\end{equation}
The expected value of $\beta$ is 
of order $1$
but the precise choice is not crucial.

To obtain the simplified black hole model, the following assumptions were made. First, the number of variables was reduced by assuming spherical symmetry. Second, instead of all possible closed graphs, a regular lattice with edge-lengths $\delta_b$ and $\delta_c$ in units of $R_P$ was used. The dynamical solution inside the event horizon
(where space is homogeneous but not static) is then obtained and corresponds to a Kantowski-Sachs solution. An analytic continuation to the region outside the horizon shows that one can reduce the two free parameters by identifying the minimum area in the solution with the minimum area of LQG.
The remaining unknown constant of the model, $\delta_b$,  is the dimensionless polymeric
parameter. Together with $A_{\rm min}$, this determines the deviation from the classical theory and must be constrained by experiment. 

The procedure to obtain the metric is as follows:
\begin{enumerate}
\renewcommand{\theenumi}{\roman{enumi}}
\renewcommand{\labelenumi}{\theenumi}
\item We define the Hamiltonian constraint, replacing the homogeneous connection with 
the holonomies along the fixed graph identified above. The diffeomorphism constraint is identically zero 
because of homogeneity and the Gauss constraint is also zero for Kantowski-Sachs spacetime. 
\item We solve the Hamilton equation of motion for the holonomic Hamiltonian system
by requiring the Hamiltonian constraint to be zero. 
\item We extend the solution to all spacetime. This is certainly legitimate mathematically but we have only found the solution in the homogeneous region, so this may not be the correct polymerization of the Hamiltonian constraint in the full spacetime.
However, we believe such a polymerization exists. 
\end{enumerate}

We now consider the form of the metric in more detail. 
Relabelling  $\delta_b$ as $\delta$, 
the metric depends only on the combined dimensionless parameter $\epsilon \equiv \delta \gamma$.
If the quantum gravitational corrections become relevant only when the curvature is in the Planckian regime, as seems plausible, then one requires $\epsilon  < 1$. 
(A stronger bound can be placed on $\epsilon$ in the solar system to avoid
excessive deviations from the classical Schwarzschild metric.)
 For $\epsilon \ll 1$, the corrections to the Schwarzschild metric outside the horizon are  of order $\epsilon^2$.
More precisely, the quantum gravitationally corrected
metric can be expressed in the form
\begin{eqnarray}
ds^2 = - G(r) c^2 dt^2 + \frac{dr^2}{F(r)} + H(r) d\Omega^{(2)},
\label{g}
\end{eqnarray}
with $d \Omega^{(2)} = d \theta^2 + \sin^2 \theta d \phi^2$ and
\begin{eqnarray}
&& G(r) = \frac{(r-r_+)(r-r_-)(r+ r_{*})^2}{r^4 +a_o^2}~ , \nonumber \\
&& F(r) = \frac{(r-r_+)(r-r_-) r^4}{(r+ r_{*})^2 (r^4 +a_o^2)} ~, \nonumber \\
&& H(r) = r^2 + \frac{a_o^2}{r^2}~ .
\label{statgmunu}
\end{eqnarray}
Here $r_+ = 2Gm/c^2$ and $r_-= 2G m P^2/c^2$ are the outer and inner horizons, respectively, and $r_* \equiv \sqrt{r_+ r_-} = 2mP$, where $m$ is the black hole mass and 
\begin{equation}
P \equiv  \frac {\sqrt{1+\epsilon^2} -1}{\sqrt{1+\epsilon^2} +1}  
\end{equation}
is the polymeric function. 
For $\epsilon \ll 1$, we have $P \approx \epsilon^2 /4 \ll 1$, so $r_-$ and $r_*$ are both much less than $r_+$. For $\epsilon \gg 1$, $P \approx 1$ and there are significant modifications to the metric beyond the horizon. Indeed, in the limit $\epsilon \rightarrow \infty$, the inner and outer horizons merge, so the black hole has no interior and one passes instantaneously between the asymptotic regions. 
Although this is a vacuum solution, we note that LQG generates an effective stress-energy tensor $T^{\mu \nu}$. 


Since $g_{\theta \theta}$ is not exactly $r^2$ in the above metric, $r$ is only the usual radial
coordinate asymptotically.  However, this coordinate
reveals the properties
of the metric most clearly. In particular, in the limit $r \to \infty$ one has 
\beqn
G(r) &\to& 1-\frac{2 G M}{c^2 r} (1 - \epsilon^2)~, \nonumber  \\ 
F(r) &\to& 1-\frac{2 G M}{c^2 r}~ , \nonumber \\
H(r) &\to& r^2 \,,
\eeqn
so the deviations from the Schwarzschild solution are of
order $G M \epsilon^2/(c^2r)$.
Here $M = m (1+P)^2$ is the ADM mass, which is determined solely
by the metric at flat asymptotic infinity and might be associated with the quantity $M$ appearing in our earlier discussion. In this section we always refer to the quantity $m$.

The expression for $H(r)$ shows that the more physical radial coordinate is 
\begin{eqnarray}
R = \sqrt{r^2 + \frac{a_o^2}{r^2}} 
\label{rho}
\end{eqnarray}
in the sense that this measures the proper circumferential distance. 
As $r$ decreases from $\infty$ to $0$, $R$ first decreases from $\infty$ to $\sqrt{2a_0}$ at $r=\sqrt{a_0}$ and then increases again to $\infty$. 
In particular, the value of $R$ associated with the event horizon is
\begin{eqnarray}
R_{EH} = \sqrt{H(r_+)} = \sqrt{ \left( \frac{2 G m}{c^2} \right)^2 + \left( \frac{a_o c^2}{2 G m} \right)^2 }\,.
\label{loopEH}
\end{eqnarray}
This is equivalent to Eq.~\eqref{quad2}, asymptoting to the Schwarzschild radius for $m \gg M_P$ and to the Compton wavelength 
for $m \ll M_P$, if we put $\xi = a_0/R_P^2 = \sqrt{3} \gamma \beta /2$. 
  
\if
Note that the quantity $R$ is a measure of the proper distance in the transverse direction. In the radial direction, the differential relationship between the proper distance $dr/F(r)$ and $dr$ is more complicated but it has the asymptotic form
\begin{equation}
\frac{\Delta \tilde r}{\Delta r} \approx
\begin{cases}
1 & (r \gg r_P) \\
(r/R_P)^{-2} & (r \ll r_P) \, .
\end{cases}
\label{radial}
\end{equation}
In particular, this result will be shown to apply in LGQ in Sec.~\ref{metric}. However, there are some ranges of $r$ in which the radial and transverse proper distances have a different dependence on $r$, so the uncertainty in position $\Delta x$ is in principle direction-dependent.
\fi

If one introduces the new coordinates  $\tilde r = a_o/r$ and 
$\tilde t = t \, r_*^{2}/a_o$, with
$\tilde r_\pm = a_o/r_\mp$ and $\tilde r_* = a_o/r_*$, the metric preserves its form.
The spacetime thus exhibits a
striking {\em self-duality} with dual radius $r=\tilde r = \sqrt{a_o}$. From 
Eq.~\eqref{rho}, this dual radius corresponds to the minimal possible
surface element and has an area $8 \pi a_o$. Since we can write Eq.~\eqref{rho} in the form 
$R = \sqrt{r^2 + \tilde r ^2} $,
it is clear that the solution
contains another asymptotically flat Schwarzschild region rather than a singularity in the limit $r\to 0$.
As is evident from the Penrose diagram in Fig.~\ref{penroseLBH}, this corresponds to a Planck-sized wormhole, whose throat is described by the Kantowski-Sachs solution. The mass of the black hole in the dual metric is 
\begin{equation}
m_{\rm ADM}^d = \frac{a_0 (1 + P)^2c^4}{4 m G^2 P^2} \, ,
\end{equation}
which is of order $M_P^2/m$ apart from the $P$ factors.

\begin{figure}
 \begin{center}
 \includegraphics[height=6cm]{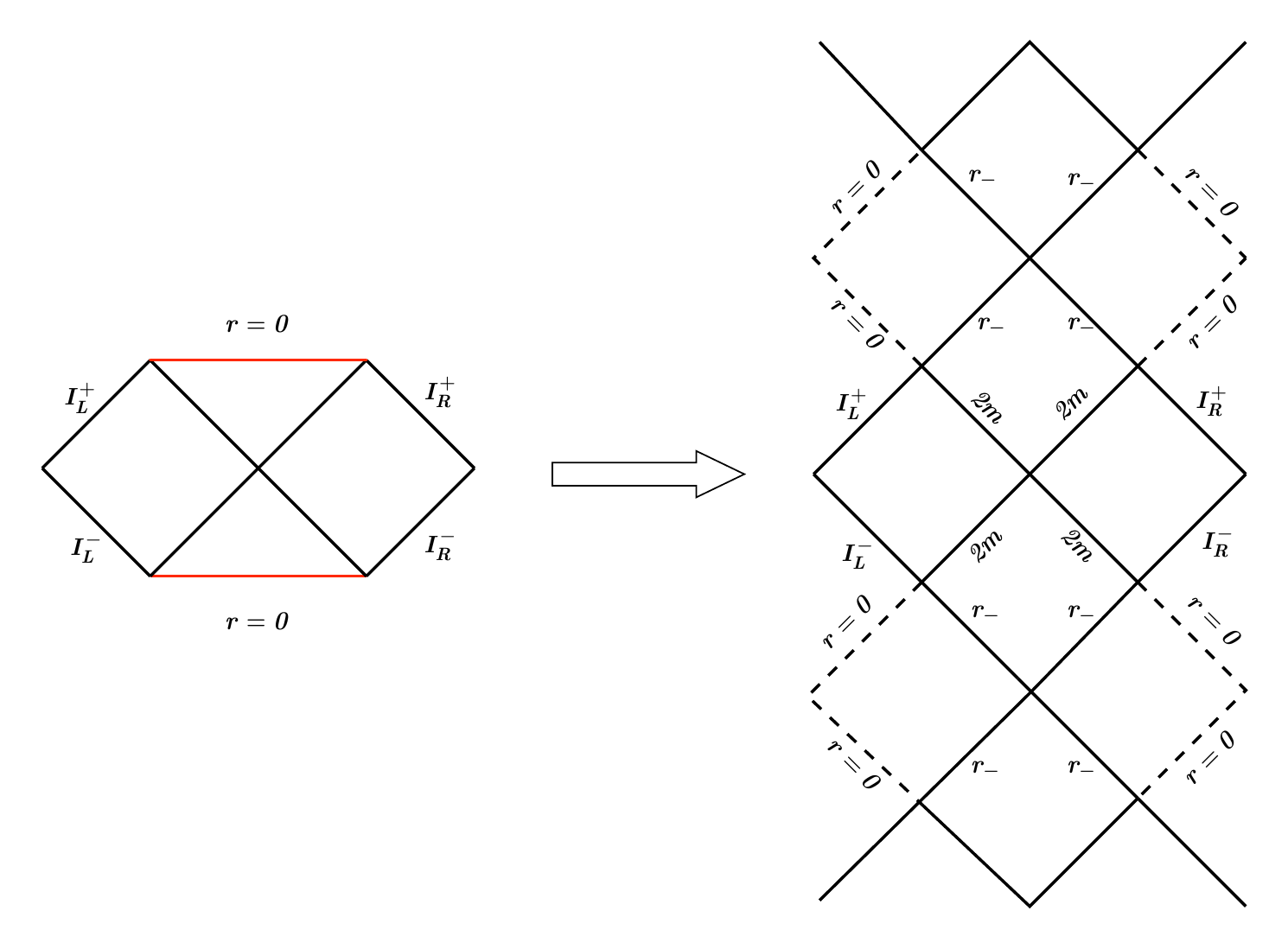}
   \end{center}
  \caption{ The Penrose diagrams for the Schwarzschild metric on the left and the LBH metric on the right. The latter has two asymptotic regions,
  one at infinity and the other near the origin, which no observer can reach in a finite time.} \label{penroseLBH}
  \end{figure}

The crucial issue is whether the event horizon is inside or outside the wormhole throat or, equivalently, whether the black hole forms before or after the bounce.  
For the event horizon to be outside the throat, we need $r_+ > \sqrt{a_o}$, which implies $m > 
(\sqrt{3} \gamma \beta /2)^{1/2} M_P$. Thus the bounce occurs {\it after} black hole formation for a super-Planckian LBH and the exterior 
is then qualitatively similar to that of a
Schwarzschild black hole of the same mass. This is illustrated in Fig.~\ref{biglqgbh}, which compares the embedding diagram for the two cases if the black hole has a mass $m = 400 \pi \sqrt{a_0}$.
The metric outside the event horizon
 differs from Schwarzschild only
by Planck-scale corrections. On the other hand, the event horizon is the other side of the wormhole throat for a sub-Planckian LBH
and the departure from the Schwarzschild metric 
is then very significant. In this case, the bounce occurs {\it before} the event horizon forms.
Consequentially, even if the
horizon is quite large (which it will be for $m \ll m_P$), it will be invisible to observers at 
$r>\sqrt{a_o}$.

\begin{figure}
 \begin{center}
 \includegraphics[height=2.5cm]{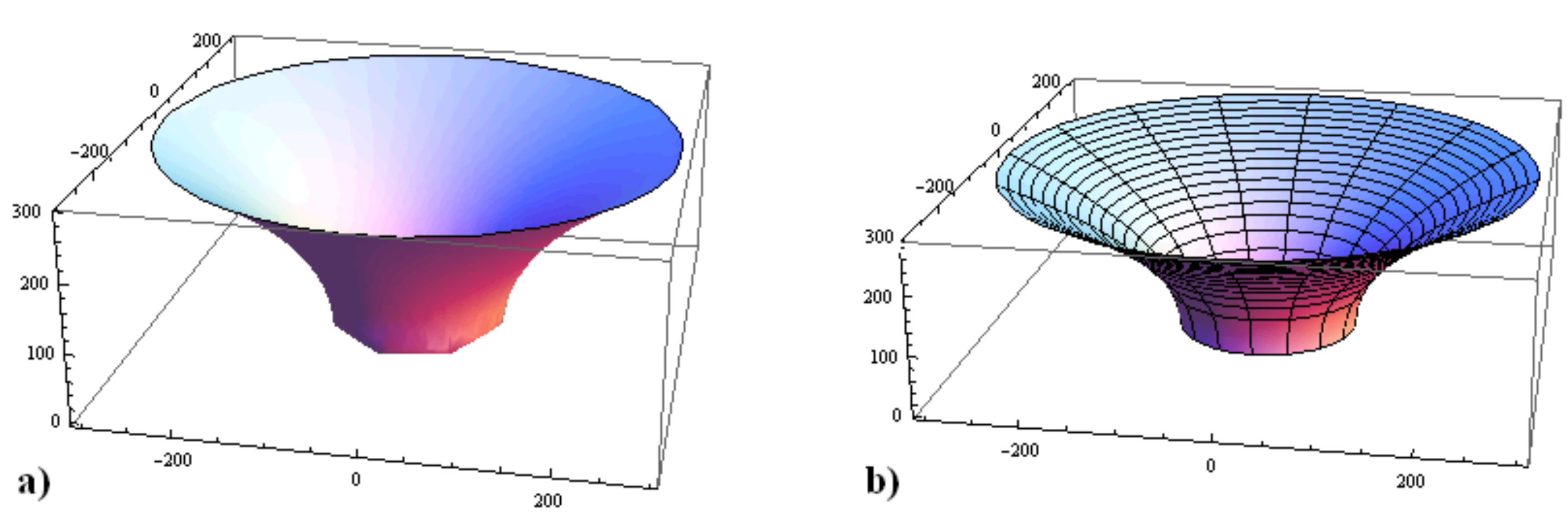}
   \end{center}
\caption{Embedding diagram of a spatial slice just outside the horizon of a $50$ Planck mass 
black hole. In (a) we have the LBH with metric (\ref{statgmunu}); (b) is the Schwarzschild black hole. In both cases the foliation is with respect to the timelike Killing vector and the scales are in Planck units. The lowest point in each diagram corresponds to the event horizon.}
\label{biglqgbh}
\end{figure}

\begin{figure}
 \begin{center}
 \includegraphics[height=2.5cm]{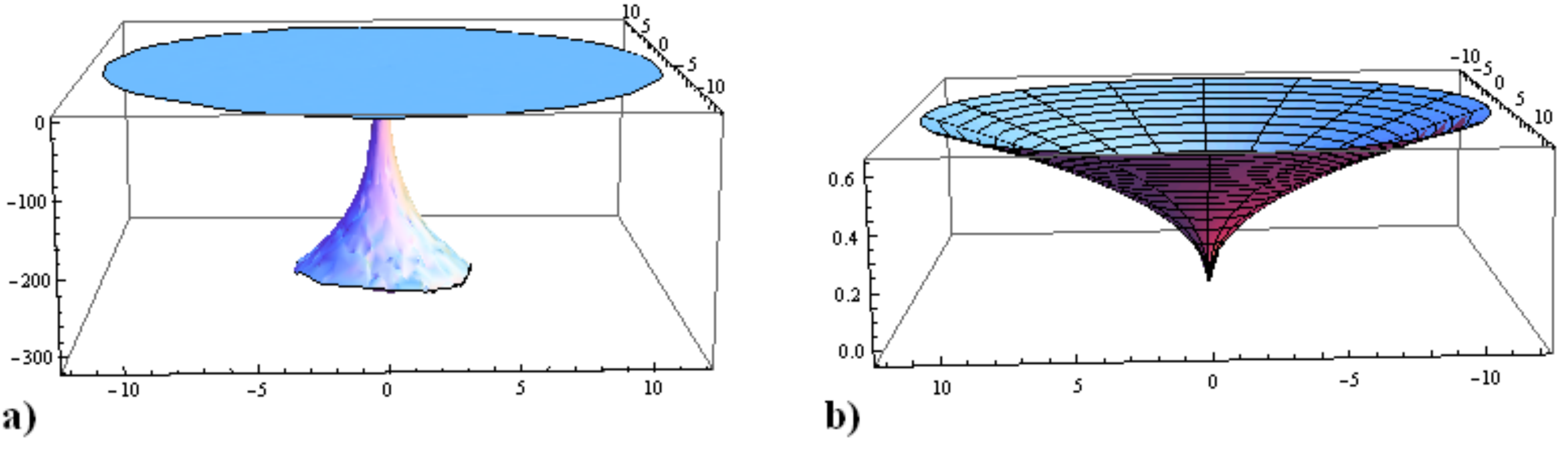}
   \end{center}
   \caption{Embedding diagram of a spatial slice just outside the horizon of a 0.005 Planck mass 
   black hole. (a) is the LBH with metric (\ref{statgmunu}); (b) is the Schwarzschild black hole. 
The foliation is the same as in Fig.~\ref{biglqgbh}
and the lowest points in each diagram correspond to the outer horizon.}
\label{smalllqgbh}
\end{figure}


It is interesting to consider the formation of these objects.  To create a black hole, we must put a mass $m$ 
inside a sphere of area $4\pi R_{EH}^2$. For $m \gg M_P$,
this area is $16\pi G^2 m^2/c^4$, so
a density $\rho_{BH} \propto m^{-2}$ is required. 
However, for $m \ll M_P$, 
we must merely localize the mass and energy 
inside a sphere of area $8\pi a_o $.
Because $A_{\rm min}\geq 5 R_P^2$ 
for the currently accepted value of the Immirzi parameter, there does not seem to be any semiclassical impediment to doing that. Hence it should be possible to create ultralight black holes.
We show the embedding diagram for
a LBH of mass $m=4\pi \sqrt{a_o}/100$ in Fig.~\ref{smalllqgbh}a.
This is contrasted with the embedding diagram of a Schwarzschild black hole of the same mass in
Fig.~\ref{smalllqgbh}b.


The most striking feature of the LBH is that the event horizon size given by Eq.~\eqref{loopEH} corresponds to the GEH expression given by Eq.~\eqref{quad2}.  This is remarkable since this result comes from geometrical rather than quantum considerations and it seems to support the $n=2$ version of the BHUP correspondence. 
This size presumably corresponds to the positional uncertainty $(\Delta x)_{BH}$ associated with the black hole. 

Let us now relate the GUP to the thermodynamic properties of black holes.
\if
we must distinguish between emitted particles and the black hole itself. 
With the usual form of the Uncertainty Principle, we obtain the black hole temperature by simply
identifying $\Delta x$ with the Schwarzschid radius and $(\Delta p)_{em}$ with some multiple $\alpha$ of the temperature:
\begin{eqnarray}
kT = \alpha c ( \Delta p)_{em} \approx \frac{ \alpha \hbar c}{\Delta x} \approx \frac{\alpha \hbar c^3}{ 2 G m} \, . 
\label{hawking}
\end{eqnarray}
For $m \gg m_P$, this gives the precise Hawking temperature if we take $\alpha = 1/(4\pi)$. For $m \ll m_P$, Eq.~\eqref{extremes} implies
\begin{eqnarray}
kT \approx \frac{ \alpha \hbar}{\Delta x} \approx \frac{2 \alpha Gm \hbar}{ a_0} \, ,
\end{eqnarray}
so the temperature increases and then decreases as $m$ increases, reaching a maximum around the Planck mass.  It is interesting that the duality between black holes of mass $m$ and $m_P^2/m$ resembles the relationship between the mass of a black hole and the energy of the particles it emits.  
For the LBH we have another way of calculating the temperature
but this appears to give a different result.
\fi
In the LBH solution, the surface gravity of the black hole is 
\begin{eqnarray}
\kappa^2 = - g^{\mu \nu} g_{\rho \sigma} \nabla_{\mu} \chi^{\rho} \nabla_{\nu}
\chi^{\sigma}/2 , 
\end{eqnarray}
where $\chi^{\mu}=(1,0,0,0)$ is a timelike Killing vector and $\Gamma^{\mu}_{\; \nu \rho}$
are the connection coefficients.
Inserting the metric into this equation,
 we find that 
the surface gravities on the
inner and outer horizons are
\begin{eqnarray}
 && \kappa_- =  \frac{4 G^3 m^3 c^4 P^4 (1-P^2)}{16 G^4 m^4 P^8 + a_o^2c^8}, \nonumber \\
 && \kappa_+ = \frac{4 G^3 m^3 c^4 (1-P^2)}{16 G^4 m^4 + a_o^2 c^8}.
\label{kpm}
\end{eqnarray}
The black hole temperature is associated with the outer horizon and given by $k T = \kappa_+/2 \pi$.
The dependence of the temperature on mass is therefore as illustrated in Fig.~\ref{logtemp}. 
It increases and then decreases as $m$ increases, reaching a maximum around the Planck mass. This confirms the qualitative behaviour anticipated in Sec.~V.

We can now reverse the argument leading to Eq.~\eqref{temp} to obtain the corresponding GUP. However, in doing so, we must distinguish between the emitted particle and the black hole itself. The uncertainty in the momentum of the emitted particle is 
\begin{eqnarray}
( \Delta p)_{em} = 4 \pi kT/c =  \frac{(2 G m)^3 (1-P^2) \hbar c^2 }{ (2 G m)^4 + a_o^2 c^8} 
\label{loopdelp} .
\end{eqnarray}
Combining this with Eq.~\eqref{loopEH}, which we interpret as the positional uncertainty of the black hole,  then gives
\begin{eqnarray}
( \Delta p)_{em} \, (\Delta x)_{BH}  \geq \frac{(1-P^2) \hbar}{ \sqrt{1+ a_o^2/(2Gm/c^2)^4} } \, .
\end{eqnarray}
For $m \gg M_P$, Eq.~\eqref{strange} implies $( \Delta x)_{BH} \approx (\Delta x)_{em}$, so we have
\begin{eqnarray}
( \Delta p)_{em} \, (\Delta x)_{em}  \geq \hbar (1-P^2)
 \left( 1-  \frac {a_o^2}{2(2Gm/c^2)^4} \right)  \, .
\label{GUPLQG2}
 \end{eqnarray}
\if
This would then imply the following GUP:
\begin{eqnarray}
 (\Delta p \Delta x )^2 \approx (1-P^2)^2 \left[1-\frac{a_o^2 (1-P^2)^2}{(\Delta x)^6 (\Delta p)^2}\right], \label{GUPLQG1}
\end{eqnarray}
or 
\begin{eqnarray}
( \Delta p)_{em} \Delta x  \geq \hbar (1-P^2) \sqrt{1-\frac{a_o^2 (1-P^2)^2}{(\Delta x)^6 (\Delta p)^2}}. \label{GUPLQG2}
\end{eqnarray}
\fi
The last term is of order $(M_P/m)^4$, which resembles Eq.~\eqref{quadapprox}. Its negative sign corresponds to a reduction in the uncertainty and is reminiscent of Eq.~\eqref{hossain} from  Hossain {\it et al.} \cite{hossain}. This term 
can be neglected for $m \gg M_P$
and observations then enable one to put a bound on $P$. 
For $m \ll M_P$, Eq.~\eqref{strange} implies $( \Delta x)_{BH} \approx (\Delta x)_{em}(m/M_P)^2$, so 
we have
\begin{eqnarray}
( \Delta p)_{em} (\Delta x)_{em}  
\geq  \frac{8}{\sqrt{3} \gamma \beta} (1-P^2) \hbar  \, .
 \label{GUPLQG4}
  \end{eqnarray}
In both cases, one obtains the standard HUP apart from numerical factors, as expected since the emitted particles have sub-Planckian energy for all values of $M$. 

This result should not be confused with the GUP for the black hole itself. Since $(\Delta p)_{BH} = mc$ and $(\Delta x)_{BH}$ is given by  Eq.~\eqref{loopEH}, 
one has  
\begin{eqnarray}
( \Delta p)_{BH} (\Delta x)_{BH}  \geq { \sqrt{3} \gamma \beta \over 4}  \hbar \sqrt{1 + {16 \over 3 \gamma^2 \beta^2} \left( { m \over M_P} \right)^4}  \, .
 \label{GUPLQG41}
  \end{eqnarray}
Apart from numerical factors, this just corresponds to the boundary shown in Fig.\ref{MR}, with $( \Delta x)_{BH} \propto (\Delta p)_{BH}^{-1}$ for $m \ll M_P$ and $( \Delta x)_{BH} \propto (\Delta p)_{BH}$ for $m \gg M_P$.

\if
Seeing as in the case of the LQBH we already have a temperature that should include both gravitational and quantum effects, we can reverse the above logic in order to obtain the corresponding GUP.

\begin{eqnarray}
 \Delta p = {4 \pi} T =  \frac{(2 G m)^3 (1-P^2)}{ (2 G m)^4 + a_o^2} \label{loopdelp}
\end{eqnarray}
and the position uncertainty, being the physical radius of the black hole is
\begin{eqnarray}
 \Delta x = \sqrt{H(r_+)} = \sqrt{(2 G m)^2 + \frac{a_o^2}{(2 G m)^2}} . \label{loopdelx}
\end{eqnarray}

Equations \eqref{loopdelx} and \eqref{loopdelp} express the GUP for the emitted particle parametrically in terms of the black hole mass.
by the following system (the mass plays the role of parameter):
\begin{eqnarray}
\begin{cases}
(\Delta p )_{em} =   \frac{(2 G m)^3 (1-P^2)}{ (2 G m)^4 + a_o^2}, \\
 \Delta x = \sqrt{(2 G m)^2 + \frac{a_o^2}{(2 G m)^2}} \, .
 \end{cases}
\label{GUPLQG1}
\end{eqnarray}

From this we can extract the functional dependence of  $\Delta x_{BH}$ upon $(\Delta p)_{em}$;
plots of $\Delta x$ versus $(\Delta p)_{em}$ and $(\Delta p)_{em}$ versus $\Delta x$
are shown in Fig.\ref{modesto2}. 
\begin{figure}
 \begin{center}
  \includegraphics[height=4.2cm]{GUPDxDp4.eps}
    \includegraphics[height=4.25cm]{GUPDpDx.eps}
  \hspace{1cm}
  \end{center}
  \caption{\label{modesto2} These graphs represent Eqs.~\eqref{loopdelx} and \eqref{loopdelp} for $(\Delta x)_{BH}$ versus $(\Delta p)_{em}$ and $(\Delta p)_{em}$ versus $\Delta x_{BH}$. The upper dashed line corresponds to the case $(\Delta x)_{BH} (\Delta p)_{em} =1$.
  We used units $a_o =1$, $\hbar =1$. The lower dashed line represents  $(\Delta x)_{BH}^3 \, (\Delta p)_{em}  =1$.
  }
  \end{figure}
We can understand the qualitative forms of this plot from Eqs.~\eqref{loopdelp} and \eqref{loopdelx} since these give
\begin{eqnarray}
(\Delta p)_{em} \propto 
\begin{cases}
(\Delta x)_{BH}^{-1} \quad (m \gg m_P) \\
 (\Delta x)_{BH}^{-3}  \quad (m \ll m_P) \, .
\end{cases}
\end{eqnarray}
By making the usual transformation $\Delta x \rightarrow R$ and $\Delta p \rightarrow mc$, the second relation corresponds to
 \begin{equation}
  R \approx (m/M_P)^{-1/3}R_P \, ,
  \end{equation} 
which is the reflection of the Planck density line in Fig.~\ref{MR}. The physical significance of this is unclear but it extends the striking symmetry between the macroscopic and microscopic domains.
\fi

\section{Discussion}

The proposal advocated in this paper leaves several unresolved questions 
 and we conclude by discussing some of these. 
One important issue within LQG is why the parameters $\alpha$ and $\beta$ appearing in the GUP and GEH expressions, respectively, should have the values required to give the exact Schwarzschild and Compton scales. This does not seem to be required {\it a priori} but has to be imposed. This is more of a problem for $\alpha$ than $\beta$, since one could argue that the coefficient in the expression for the Compton scale is somewhat arbitrary anyway. 

Another puzzle concerns the relationship between sub-Planckian black holes and elementary particles in LQG, since they both
lie on the Compton line in Fig.~\ref{MR}. One distinction is that  
the quantity $R$ is measured in different asymptotic spaces in the two cases. This is more apparent if one represents Fig.~\ref{MR} in terms of the (unphysical) coordinate $r$ rather than $R$. The black hole and elementary particle lines are then clearly separated, although $r$ becomes sub-Planckian in the black hole asymptotic space. In any case, there is clearly some deep connection here. This is described as Quantum Particle Black Hole Duality and discussed further elsewhere \cite{poly}. 

It is also unclear is how the {\it formation} of a LBH is to be represented in Fig.~\ref{MR}. 
For a light LBH ($M \ll M_P$), the collapse of a spherically symmetric region is presumably represented by a downwards vertical trajectory. This first goes below the Compton line and then reverses direction when the region bounces (at sub-Planckian density) to form a Planckian wormhole. 
The region then re-expands into another asymptotic region and appears as a sub-Planckian white hole when it re-crosses the Compton line. However, a process in which a region falls below its Compton wavelength is problematic and it is possible that a sub-Planckian black hole could only form via the evaporation of a super-Planckian one. 
The formation of a heavy LBH ($M \gg M_P$) is equally problematic because the collapsing region crosses the Planck density line before bouncing at the Planck volume.  
Yet another problem is that the duality between super-Planckian and sub-Planckian black holes suggests that they should form together in the LQG scenario and this is difficult to represent in Fig.~\ref{MR}. 
We discuss these issues in a separate paper~\cite{cmp2}.

Moving to more general quantum gravity scenarios, some link between the Uncertainty Principle and black holes is  suggested by the work of Calmet {\it et al.} \cite{calmet}. They argue that the Planck scale $R_P$ must be the minimal length on the  basis of causality, the Uncertainty Principle and a dynamical criterion for gravitational collapse called the ``hoop conjecture''. 
There may also be some link with the notion of fuzzy black holes \cite{silva} and non-cummutative black black holes \cite{majid}. 

Unfortunately, it seems unlikely that the BHUP correspondence applies
in the context of string theory, which is the other main contender for a theory of quantum gravity.
While Eq.~\eqref{GUPstring} formally resembles Eq.~\eqref{GUP1}, so that some form of GUP seems to apply, we have seen that the second term cannot correspond to a black hole for $M \gg M_P$ 
because  the string is too elongated to form an horizon. 
The situation is even worse for $M < M_P$ because Maggiore \cite{maggiore} argues that the concept of a black hole is then not operationally defined. 
A formula like Eq.~\eqref{newR} also arises in string theory, with the two terms  corresponding to the string excitation (which scales as $r^{-1}$) and 
the string winding (which scales as $r$) but the energies add linearly rather than quadratically. 

 Another important perspective comes from Giddings and Lippert \cite{giddings2}, who argue that the Schwarzschild line in Fig.~\ref{MR} represents the boundary of the domain of validity of local quantum field theory (QFT). 
 The Uncertainty Principle implies that the phase-space description 
 of classical mechanics must break down somewhere, so we need to parameterize for what field configurations QFT is no longer valid.  
 Their approach uses
 a perturbative Fock space description of QFT states, in which the 
incorporation of gravity is best represented in the two-particle sector. However, Fock space states that fall below the Schwarzschild line do not have a perturbative description in simple QFT terms. This is described as a ``locality bound'' and it  specifies the boundary in Fig.~\ref{MR}
where QFT must match onto a more complete theory of quantum gravity.  Taking this to be
 the Schwarzschild line also
has the advantage that it can be reconciled with Lorentz-invariance. A diagram similar to Fig.~\ref{MR} arises if one thinks of scattering as a function of center-of-mass energy and impact parameter~\cite{giddings3}. 

The principle of locality is clearly very relevant to the considerations of this paper. Doplicher~\cite{doplicher} points out that this is a crucial feature of combining special relativity with quantum theory and it is associated with a {\it global} gauge-invariance. However, this no longer applies when one accounts for the gravitational interaction between particles, so there is no {\it local} gauge-invariance, and this is why 
the merger of general relativity with quantum theory removes locality on much larger scales than the Planck length. The BHUP correspondence suggests that it breaks down at the black hole scale.

The Uncertainty Principle itself acquires a different significance in string theory, being interpreted in terms of the 4-dimensional relationship
$\Delta t  \, \Delta x > R_P^4$.
This has been discussed by Yoneya \cite{yoneya1}, who gives a semi-classical reformulation of string quantum mechanics in which the dynamics is represented by the non-commutativity between temporal and spatial coordinates.
More recently he has extended this to field theories involving D-branes  and black holes \cite{yoneya2}.

It is particularly interesting to extend the BHUP correspondence to the higher dimensional case. It is clear that new physics is encountered at the Planck scale, with the minimum length-scale corresponding to a UV cut-off \cite{hossenfelder}. However, in some versions of string theory there are large extra dimensions and in this case the Planck scale is lowered, possibly into the experimentally accessible range. This might allow the production of higher-dimensional black holes in accelerators.  
Since we  know how the scaling of the black hole radius with $M$ changes with dimensionality in most models, one should be able to predict the form of the GUP explicitly. 


Finally, we note that there is an interesting contrast between the LBH properties and those of the black holes which arise in ``asymptotically safe'' gravity~\cite{falls}. In these  models, gravity weakens at small scales, so that the event horizon is smaller than usual. This implies that the black hole boundary in Fig.~\ref{MR} becomes steeper as $M$ decreases, with the horizon size going to zero at some critical mass.  As with the LBH, a central singularity is avoided but without invoking another asymptotic region. Nevertheless, the metric and Penrose diagram have a similar form to the LBH case and one again has  
Planck-size remnants with vanishing temperature.

\if
One important aspect of the BHUP correspondence is its implications for black hole temperature and entropy.
 Quantum corrections to the Bekenstein-Hawking entropy formula usually introduce a logarithmic term, described by the Cardy-Verlinde formula, but we have seen that the BHUP correspondence implies another type of correction.
There is also the suggestion that the black hole surface area has a discrete spectrum, with a uniform spacing determined by the area-entropy relationship and the Boltzmann-Einstein distribution \cite{hod}. However, it is not clear how this relates to the present proposal.

A black hole with $m \gg M_P$ has $T \propto m^{-1}$ in this universe but it is also associated with a black of mass $\mu \propto m^{-1}$ and hence $T \propto \mu^3 \propto m^{-3}$ in the other universe.
A black hole with $m \ll M_P$ has $T \propto m^{3}$ in this universe but it is also associated with a black of mass $\mu \propto m^{-1}$ and hence  $T \propto \mu^{-1} \propto m$ in the other universe.
So one has four different power laws.

Another unsatisfactory feature of Fig.~\ref{MR} is that the overlap between the quantum and relativistic regimes - in some sense the holy grail of modern physics -- appears to be the single point given by  Eq.~\eqref{planck}, although we have seen that quantum gravitational effects may extend into the entire region $\rho > \rho_P$.  Our other paper examines two ways of overcoming this problem: one invokes extra dimensions; the other involves extending Fig.~\ref{MR} into sub-Planckian scales.
\fi

\section*{Acknowledgements}

We are grateful to Ronald Adler, Xavier Calmet, Steve Giddings, Craig Hogan, Bernard Kay and Rodolfo Russo for helpful discussions. BJC thanks the Canadian Institute for Theoretical Astrophysics, the Perimeter Institute, the Astrophysics Group at Fermilab and the Department  of Mathematics and Statistics at Dalhousie University for hospitality received during this work. Perimeter Institute is supported by the Government of Canada through Industry Canada
and by the Province of Ontario through the Ministry of Research \& Innovation. This work
was supported in part by the Natural Sciences \& Engineering Research Council of Canada, the Humboldt Stiftung and the Science and Technology Facilities Council of the UK.


\vspace*{0.5cm}

\begin{thebibliography}{99}

\bibitem{carr-rees}
B J Carr and M J Rees, 
Nature 278, 605 (1979).

\bibitem{adler1}
R J Adler, 
[arXiv:1001.1205 [gr-qc]].

\bibitem{hawking}
S W Hawking, Nature 248: 30 (1974); Comm. Math. Phys. 43:199 (1975).

\bibitem{Modesto:2008im}
  L Modesto 
   Int. J. Theor. Phys 2010
  [arXiv:0811.2196 [gr-qc]]

\bibitem{bowick}
M J Bowick et al., Phys. Rev. Lett. 61, 2823 (1988); S Coleman, J Preskill \& F Wilczek, Mod. Phys. Lett. A. 6, 1631 (1991).

\bibitem{heisenberg}
W Heisenberg, 
Zeitschrift fur Physik 43, 172198 (1927). 

\bibitem{kennard}
E H Kennard, 
Zeitschrift fr Physik 44, 326 (1927). 

\bibitem{oppenheim}
J Oppenheim and S. Wehner, Science 330, 1072-1074 (2010).


\bibitem{wheeler}
J A Wheeler,  Geons, Phys. Rev. 97, 511-536 (1955)


\bibitem{Adler}
R J Adler and D I Santiago, 
Mod. Phys. Lett. A14, 1371 (1999) [gr-qc/9904026]; 
R J Adler, P Chen and D I Santiago, 
Gen. Rel. Grav. 33, 2101 (2001); 
P Chen and R J Adler, 
[gr-qc/0205106]; 
P Chen, 
[arXiv:astro-ph/0303349].


\bibitem{chang}
L N Chang et al., Phys. Rev. D 65, 125028 (2002); 
R Zhao, Y Q Wu and L C Zhang Class. Quant. Grav. 20, 4885 (2003); X. Li, Phys. Lett. B. 540, 9 (2002).


\bibitem{das}
S Das and E C Vagenas, Phys. Rev. Lett. {\bf 101}, 221301 (2008);  S Das and E C Vagenas, 
Ca. J. Phys.  87, 233-240 (2009).


\bibitem{mead}
C A Mead, Phys. Rev. 135, 849 (1964).

\bibitem{maggiore}
M Maggiore, Phys. Lett. B 304, 65 (1993); Phys. Lett. B 319, 83 (1993); 
M Maggiore, Phys. Rev. D. 49, 5182 (1994).

\bibitem{veneziano}
G Veneziano, Europhys.Lett. 2, 199 (1986); 
D J Gross, 
PUPT-1108 Plenary Session, 24th Int. Conf. on High Energy Physics, Munich, Aug 4-10 (1988);
E Witten, Phys. Today April 24 (1996);
F Scardigli, Phys. Lett. B 452, 39 (1999); D J Gross and P F Mende, Nuc.Phys.B303, 407 (1988); 
D Amati, M Ciafaloni and G Veneziano, Phys, Lett. B. 216, 41 (1989); 
T Yoneya, Mod. Phys. Lett. A4, 1587 (1989); K Konishi, G Paffuti and P Proverpo, 
Phys. Lett. B 234, 276 (1990). 


\bibitem{ashtekar}
A Ashtekar, S Fiarhurst and J L Willis, Class. Quant. Grav. 20, 1031 (2003).

\bibitem{hossain}
G M Hossain, V. Husain and S.S. Seahra, [arXiv:1003.22071 [gr-qc]].

\bibitem{kay}
B S Kay, 
Class. Quant. Grav.15, L89-L98 (1998);
B S Kay and V Abyaneh, 
[arXiv:0710.0992].




\bibitem{camellia}
P S Custodio, Class. Quant. Grav. 20, L197-L203 (2003); 
G A Camellia, M Arzano and A Procaccini, Phys. Rev. D 70, 107501 (2004); 
M R Setare, Phys. Rev. D 70, 087501 (2004); Int. J. Mod. Phys. A 21, 1325 (2006); 
Z Ren and Z Sheng-Li, Phys. Lett. B. 641, 208-211 (2006).


\bibitem{hod}
S Hod, Phys. Rev. Lett. 81, 4293-4296 (1998).

\bibitem{LQGgeneral}
C Rovelli, {\sl Quantum Gravity}, Cambridge University Press,
Cambridge (2004); 
A Ashtekar, 
Class.\ Quant.\ Grav.\ 21, R53 (2004) [arxiv:gr-qc/0404018];
T Thiemann, 
[hep-th/0608210];
[gr-qc/0110034];
T.~ Thiemann, 
Lect.\ Notes Phys.\ 631, 41-135 (2003) [arxiv: gr-qc/0210094].

\bibitem{Bojowald}
M Bojowald,  
Living Rev. Rel. 8, 11 (2005)
[gr-qc/0601085];
A Ashtekar, M Bojowald and J Lewandowski,
Adv. Theor. Math. Phys. 7,  233-268 (2003) [gr-qc/0304074];
M Bojowald,
Phys. Rev. Lett. 86, 5227-5230 (2001)
[gr-qc/0102069].



\bibitem{poly}
L Modesto, I Premont-Schwarz, 
Phys. Rev. D 80, 064041 (2009)
[arXiv:0905.3170 [hep-th]];
L Modesto,
Adv. High Energy Phys. 2008, 459290 (2008)
[gr-qc/0611043];
L Modesto,
Class. Quant. Grav. 23, 5587-5602 (2006)
[gr-qc/0509078];
A Ashtekar and M Bojowald,
Class. Quant. Grav. 23, 391-411 (2006) 
[gr-qc/0509075];
L Modesto, 
Class. Quant. Grav. 23, 5587-5602 (2006),
[gr-qc/0509078];
L Modesto,
Phys. Rev. D 70, 124009 (2004)
[gr-qc/0407097];
L Modesto,
Int. J. Theor. Phys. 45, 2235-2246 (2006) 
[ arXiv:gr-qc/0411032];
C G. Bohmer, K. Vandersloot,
arXiv:0709.2129;
D W Chiou,
Phys. Rev. D 78, 064040 (2008)
[arXiv:0807.0665];
J Ziprick and G Kunstatter, 
[arXiv:1004.0525 [gr-qc]];
G Kunstatter, J Louko, A Peltola,
Phys. Rev. D 81, 024034 (2010)
[arXiv:0910.3625 [gr-qc]];
J Ziprick and G Kunstatter,
Phys. Rev. D80, 024032 (2009)
[arXiv:0902.3224 [gr-qc]];
A Peltola, G Kunstatter,
Phys. Rev. D 80, 044031 (2009)
[arXiv:0902.1746 [gr-qc]];
F Caravelli, L Modesto,
Class. Quant. Grav. 27, 245022 (2010)
[arXiv:1006.0232];
L Modesto, J W Moffat, P Nicolini, 
Phys. Lett. B 695, 397-400 (2011)
[arXiv:1010.0680 [gr-qc]].

\bibitem{Hossenfelder:2009fc}
  S Hossenfelder, L Modesto, I Premont-Schwarz,
Phys. Rev. D81, 044036 (2010)
[arXiv:0912.1823 [gr-qc]]; 
E. Alesci, L. Modesto, 
[arXiv:1101.5792 [gr-qc]];
E Brown, R B Mann, L Modesto, 
Phys. Lett. B 695, 376-383 (2011) 
[arXiv:1006.4164 [gr-qc]];
E G Brown, R B Mann, L Modesto, 
[arXiv:1104.3126 [gr-qc]].

\bibitem{brannlund}
J. Brannlund, S. Koster and A. DeBenedictis, [arXiv:0901.0010 [gr-qc]]; 
A. DeBenedictis, 
[arXiv:0907.0826 [gr-qc]].

\bibitem{GP}
R Gambini, J Pullin, 
Phys. Rev. Lett. 101, 161301 (2008)
[arXiv:0805.1187];
M Campiglia, R Gambini, J Pullin, 
AIP Conf. Proc. 977, 52-63 (2008)
[arXiv:0712.0817];
M Campiglia, R Gambini, J Pullin, 
Class. Quant. Grav. 24, 3649-3672 (2007)
[gr-qc/0703135];
V Husain, D R Terno, 
Phys. Rev. D 81, 044039 (2010)
[arXiv:0903.1471 [gr-qc]].


\bibitem{AA} A Ashtekar,
Phys.\ Rev.\ Lett.\  57 (18): 2244-2247 (1986).

 
\bibitem{universemass}
D Valev,
[arXiv:1004.1035v1 [gen-ph]].


\bibitem{RNstability}
R Matzner {\it et al.}, 
Phys. Rev.  D 19, 2821 (1979).


\bibitem{NCBH}
 P Nicolini, A Smailagic and E Spallucci 
[arXiv:hep-th/0507226];
%
 P Nicolini, 
 J. Phys. A  38, L631 (2005)
 [arXiv:hep-th/0507266];
%
 P Nicolini, A Smailagic and E Spallucci, 
 Phys. Lett.  B 632, 547 (2006)
 [arXiv:gr-qc/0510112];
%
 S Ansoldi, P Nicolini, A Smailagic and E Spallucci, 
 Phys. Lett.  B 645, 261 (2007)
 [arXiv:gr-qc/0612035];
%
 E Spallucci, A Smailagic and P Nicolini, 
 Phys. Lett. B 670, 449 (2009)
 [arXiv:0801.3519 [hep-th]];
%
 P Nicolini and E Spallucci, 
 Class. Quant. Grav. 27, 015010 (2010)
 [arXiv:0902.4654 [gr-qc]];
%
%
 P Nicolini, M Rinaldi, 
 [arXiv:0910.2860 [hep-th]];
 M Bleicher and P Nicolini, 
 [arXiv:1001.2211 [hep-ph]];
 D Batic, P Nicolini, 
 [arXiv:1001.1158 [gr-qc]];
%
%
%
Y S Myung, Y W Kim, Young-Jai Park, 
[arXiv:0708.3145]; 
Y S Myung, Y W Kim, Y J Park, 
Phys. Lett. B  656, 221-225 (2007)
[gr-qc/0702145];
P Nicolini,
[arXiv:0807.1939];
R Banerjee, B R Majhi, S Samanta, 
Phys. Rev. D 77, 124035 (2008) 
[arXiv:0801.3583];
R Banerjee, B R Majhi, S K Modak, 
[arXiv:0802.2176];
 V Husain, R B Mann, 
Class. Quant. Grav. 26, 075010 (2009)
[arXiv:0812.0399 [gr-qc]];
A Bonanno, M Reuter,
Phys. Rev. D 62, 043008 (2000);
A Smailagic, E Spallucci
[arXiv:1003.3918]; 
 L Modesto, P Nicolini, 
 Phys. Rev. D 82, 104035 (2010)
 [arXiv:1005.5605[gr-qc]].

\bibitem{ff} 
L Brewin, 
Gen. Rel. Grav. 39, 521-528 (2007)
[gr-qc/0609079].



\bibitem{bh3}
P Nicolini,
Int. J. Mod. Phys. A 24, 1229 (2009).

\bibitem{bh4}
P Nicolini, A Smailagic and E Spallucci,
Phys. Lett. B 632, 547 (2006).





\bibitem{Hossenfelder:2009xq}
  S Hossenfelder and L Smolin,
  Phys.\ Rev.\  D 81, 064009 (2010)
  [arXiv:0901.3156 [gr-qc]].





\bibitem{poisson}
E Poisson, W Israel,
Phys. Rev. D 41, 1796 (1990).

\bibitem{poisson2}
R Balbinot, E Poisson,
Phys. Rev. Lett. 70, 13 (1993).

\bibitem{cmp2}
B J Carr, L Modesto and I Premont-Schwarz, Cosmological implications of Quantum Particle Black Hole Duality, preprint (2011).

\bibitem{calmet}
X Calmet, M Graesser and S D H Hsu, Phys. Rev. Lett. 93, 211101 (2004); 
X Calmet, M Graesser and S D H Hsu, [hep-th/0505144].

\bibitem{silva}
J Madore, 
Class. Quant. Grav. 9, 69-88 (1992); 
B P Dolan, 
JHEP 0502:008 (2005); C. A. S. Silva, 
[arXiv:0812.3171].

\bibitem{majid}
S Majid, 
[arXiv:1009.2201 [math.QA]]. 

\bibitem{giddings2}
S B Giddings and M Lippert, 
Phys. Rev. D. 65, 024006 (2002) [hep-th/0103231]; 
S B Giddings and M Lippert, 
Phys. Rev. D. 69, 124019 (2004) [hep-th/0402073].

\bibitem{giddings3}
S B Giddings, [arXiv:0910.3140 [gr-qc]];  
S B Giddings, [arXiv:0709.1107 [hep-ph]]; 
S B Giddings and R A Porto, [arXiv:0908.0004]. 
  
\bibitem{doplicher}
S Doplicher, J. Math. Phys. 51, 015218 (2010).

\bibitem{yoneya1}
T Yoneya, [hep-th/0004074]; 
T Yoneya, 
[hep-th/0010172].

\bibitem{yoneya2}
T Yoneya, Prog. Theor. Phys. Supp. 171, 87-98  (2007) [arXiv:0706.0642 [hep-th]].

\bibitem{hossenfelder}
  S Hossenfelder,
  Mod. Phys. Lett. A. 19, 2727-2744 (2004).
 
\bibitem{falls}
K Falls, D F Litim and A Raghuraman, 
[arXiv:1002.0260].

\end{thebibliography}
\end{document}